# A unified theory correcting Einstein-Laub's electrodynamics solves dilemmas in the electromagnetic stress tensors and photon momenta

M.R.C. Mahdy[1], Dongliang Gao[1], Weiqiang Ding[1], M. Q. Mehmood [1], Manuel Nieto-Vesperinas[2], Cheng-Wei Qiu[1*]

[1]Department of Electrical and Computer Engineering, National University of Singapore, 4 Engineering Drive 3, Singapore 117583

[2] Instituto de Ciencia de Materiales de Madrid, Consejo Superior de Investigaciones Cientificas, Campus de Cantoblanco, 28049, Madrid, Spain

ABSTRACT: **To unify and clarify the persistently debated electromagnetic stress tensors (ST) and photon momenta, we establish a theory inspired by the Einstein-Laub formalism inside an *arbitrary* macroscopic object immersed in any complex medium. Our generalized Einstein-Laub force and ST yield the total force experienced by any *generic* macroscopic object due to the internal field interacting with its atoms, charges and molecules. Appropriate scenarios are established for the conservation of a newly proposed momentum that we call *non-mechanical generalized Einstein-Laub momentum, along* with the kinetic and canonical momenta of photons. Our theory remains valid even in a generally heterogeneous or bounded embedding background medium without resorting to *hidden momenta,* and unambiguously identifies the 'existence domain', or validity domain, of the STs and photon momenta proposed to date. This 'existence domain' is the region either outside a macroscopic scatterer with only exterior fields, or at its interior with only inside fields. The appropriate identification of such 'existence domain' constitutes the basis of our unified theory. Finally, a thought experiment is proposed, which shows that the appropriate force and the photon momentum in the embedding medium can also be properly identified if the background is comparatively larger than the embedded scatterer. It also explains the fully different roles of the Abraham and Minkowski photon momenta in the embedding medium. Most importantly, our unified theory reveals that a unique formulation of the momentum conservation law is unfeasible, though a generalized expression of the ST and momentum density is achievable in terms of new  concepts  that we introduce, namely, the 'effective polarization' and 'effective magnetization'.**

# I. Introduction

Radiation pressure, in conjunction with photon momentum, has always constituted an intriguing phenomenon in physics [1-4]. An accurate prediction of the electromagnetic force is important for complex biological systems [5, 6], stable optical manipulation [7,8], tractor beams [7,9,10], MEMs and nano-opto-mechanical systems [11,12], among many applications. Most experiments in those areas need determinations of these optical forces [5-10,13,14] on objects immersed in a non-vacuum environment. However, inside matter different definitions and descriptions of the macroscopic optical force and the photon momentum have been put forward, among others by Minkowski [4], Abraham [4], Nelson [15], or Peierls [16]. Even after extensive debates spanning over a century [2, 3], there are still some unsolved problems regarding exactitudes of stress tensors (ST) or force laws, (cf. Table 1 in [4]), and on their individual validities or limitations inside matter. Some details on the conflicts involving the STs by Einstein-Laub (EL), Minkoswki, Chu, and Nelson can be found in [2-4, 13, 14, 17-23].

At this point we must specify that throuoghout this paper we refer to 'exterior' or 'outside' magnitudes as those evaluated outside the volume of the embedded macroscopic object, while by 'interior' or 'intside' we shall refer to those quantities inside this object volume.

Some of the most important dilemmas are listed here: (i) Minkowski ST has been used to determine the force on an object from exterior fields [2, 4, 17, 18], but such formulation was shown to be problematic for obtaining the 'outside force' in [19, 20]. Also, the STs and force laws employed in [15, 16] are different to the Minkowski ST used to obtaining the 'outside force'. (ii) If interior fields of the body are employed, Minkowski ST only yields the force from the conducting currents rather than the total force [17]. Though the total volume force on an object embedded in air or vacuum can be calculated based on the force equation presented in [17], it requires the manual addition of a hidden quantity [21]. (iii) The prediction of the Einstein-Balazs box thought experiment in favor of Abraham photon momentum $p_{Abr}$ [2] has never been directly verified with any ST formula that leads to $p_{Abr}$ while it yields the *total* interior force on an absorbing or non-absorbing scatterer. (iv) A recent study [21] supports the EL force law [22] (instead of Lorentz-Nelson's one [15]) which does not require any hidden momentum [21]. Nevertheless the applicability of the EL force of [21] has been seriously questioned in several simple experimental situations [23], (specially the problematic definition of magnetic induction ***B*** in the EL force). (v) Most importantly, splitting the photon momentum into canonical (i.e. Minkowski, $p_{Mink}$) and kinetic (i.e. Abraham, $p_{Abr}$) momenta [24], though conceptually insightful, does not constitute an operational and general procedure to determine appropriate optical forces [20] in different complex and experimental situations like those discussed in [3,7,13,14, 19,20]. Then an ultimate question may be raised: is it possible to get a unique ST, electromagnetic momentum density, and force law able to handle almost all practical situations, and previous experimental observations, in contrast with those previous ambiguous formulations listed in Table-1?

An answer, and hence a conclusioon on whether a unique linear momentum conservation equation is possible or not, is given in detail in the *Final Remarks* Section of this paper. However, the fact is that no previously reported ST, force law, and electromagnetic momentum density, (listed in Table-1 in [4]) is incorrect, (cf. Supplement S0 and S1). Rather, they all have their specific validity range, not yet established, which we henceforth characterize in this paper as '*existence domain*' (cf. also Ref. [25]) as the region either outside a scattering body taking only its exterior fields into account, or in its interior considering only the inside fields. Previous works [15-24] may have overlooked this 'existence domain' of the different macroscopic STs and associated electromagnetic momentum densities. This has hindered the consecution of a solution, such as the one put forward in this paper, which identifies the corresponding validity regions of the macroscopic STs, force equations (without hidden force) and photon momenta proposed to date. They are explained in Table-1. Further, in this paper we unify them all in a compact framework, [cf. Figs. 1 (a) and (b)]. More arguments justifying the concept: 'existence domain' are presented at the end of this paper before the *Final Remarks* Section (in section VIII and IX).

In this context we must point out that still some notable unsolved problems remain, such as:

(a) In the biological cell trapping experiment reported in [26] it is yet unexplained which force and ST are more accurate [4] if the embedding background is considered water instead of air. Though Chu [27], Einstein-Laub [4,28] and Minkowski forces [4] lead to the same result when the background is air or vacuum, it is not yet known which theory is adequate for the configuration of [26], that considers a non-vacuum embedding medium. (b) The experiment of [29] provided the first quantitative agreement between optical trap stiffness and calculations. Since it was considered that such experiment supported the EL force [4] as the 'outside force' with background fields, is then our proposal regarding the EL force in Table-1 incorrect? (c) In optical tweezer [30] set-ups, such as tractor beams on fully immersed objects [31], and biophysics experiments [32], approximate methods such as those using the dipole force [13], sometimes are not valid. For example, in [13], even being the object dipolar, the dipole force fails to be correct, (cf. Fig. 3 (e) and (f) in [13]). Is then possible to attain a generalized formulation that yields both exterior and interior forces without resorting to problematic hidden quantities, in these aformentiomed cases? (d) Most importantly, there is no generalized ST along with the electromagnetic momentum density, which correctly yields the force inside a generic object inmmersed in an arbitrary background, and that correctly accounts for: (i) The experimental observations in [13, 14, 26, 29], and (ii) Some of the predictions of [31], while they comply with our imperative of holding with the 'existence domain' requirement.

To handle and fully explain the complex problems of points (i) to (v), and (a) to (d) above, without using problematic hidden quantities, (cf. Supplement S0 and S1), we establish a unified theory of linear momentum conservation by introducing new concepts. These are: the *generalized Einstein Laub (GEL) system,* (cf. Fig. 1(c) and Fig.2) and the *GEL non-mechanical momentum* of photons. This is done based on the bridge established by our generalization of the Einstein-Laub ST, electromagnetic momentum density and the force law, which we express in terms of an *effective polarization* and an *effective magnetization* that we introduce in this work.

This generalized formulation not only supports the recent observation in [33], but also at difference with previous studies, it tackles and successfully interpretes other more complex situations such as:

(I) The interior force dynamics of a generic object, (e.g. a Rayleigh, dipole, Mie, or more complex than Mie, body), embedded in a generally heterogenous background, [cf. Fig. 2(a)], or in a bounded background [see Fig. 2(b)]. (II) The interior force in highly absorbing magnetodielectric objects, and the force dynamics of the embedding background itself. (III) By employing only the interior fields of an embedded object, our generalized EL equations validate the previous experimental observations [3,7,13,14,32-35] and theoretical predictions [36-40] of the outside force by a Minkowskian formulation, (rather than the opposite proposals in [15,16,19,20,41,42]), among all STs and electromagnetic momentum densities listed in Table-1. Specially, the observation of the conflicting momenta $p_{Abr}$ [20] and $p_{Mink}$ [3, 35] in the experiment of [34], is also explained. (IV) More importantly, *our unified theory resolves long-lasting debates between the formulations of Abraham* [3, 4, 19], *Einstein-Laub* [20-23] *and Minkowski* [3,4,17,18] *by demarcating their 'existence domains' and unifying them in a compact framework via our generalization of the EL equations.*

**II. Previous stress tensors, electromagnetic momentum densities and force laws: Their 'existence domain' without hidden quantities**

To get a clear view of the operations physically involved in the different STs, force densities and photon momenta [4] established to date, and their 'existence domain', we list them in Table-1. To establish a unified theory, it is imperative to understand the behaviors of all these quantities. Later, in order to unify them all, we define three different systems including our generalization of the Einstein-Laub equations, (cf. Figs. 1 (a)-(c) and Fig. 2).

**Table-1: Previous macroscopic tress tensors, electromagnetic momentum densities and force laws: their 'existence domain' without hidden quantities (HQ)***

| | Stress tensor, $\overline{\overline{T}}$, and force density, $f$ | Electromagnetic momentum density, $G$ | Comment |
|---|---|---|---|
| Minkowski | $\overline{\overline{T}} = DE + BH - \frac{1}{2}(B \cdot H + D \cdot E)\overline{\overline{I}}$ <br><br> $f = -\frac{1}{2}E^2\Delta\varepsilon - \frac{1}{2}H^2\Delta\mu$ | $D \times B$ | The Minkowski ST correctly chrachtarizes only the information of the transfer of momentum transported by the field from any background medium to an embedded scatterer. The time averaged Minkowski ST yields the 'outside force' only from fields ouside the embedded object. It does not require HQ*. The correct 'outside force' can also be obtained directly from calculations employing $P_{Mink}$ (i.e. by ray tracing method). The time averaged Minkowski ST yields a zero mechanical force ('felt force') at the interior of a lossless object and is not applicable inside an object. Most of the major experimental works regarding photon momentum are based on changes in space and not in time. As a result, both in the background and inside the embedded body, the observable or measurable linear photon momentum is so far Minkowski's. |
| Abraham | $\overline{\overline{T}} = \frac{1}{2}\left[DE + ED + BH + HB - (B \cdot H + D \cdot E)\overline{\overline{I}}\right]$ <br><br> $f = \left(-\frac{1}{2}E^2\Delta\varepsilon - \frac{1}{2}H^2\Delta\mu\right) + f_A;$ <br><br> $f_A = (n^2 - 1)\frac{\partial}{\partial t}\left(\frac{E \times H}{c^2}\right)$ | $\frac{E \times H}{c^2}$ | The time averaged Abraham ST yields the same 'outside force' as Minkowski ST at the exterior of an object just employing the outside fields when the embedding medium is not air or vacuum (but isotropic). However, the Abraham force law is more appropriate to characterize the 'outside force' in time domain due to the extra $f_A$ term in its force density**. The correct 'outside force' cannot be directly obtained from the $P_{Abr}$ (i.e. by ray tracing method). Inside an absorbing object, both time averaged Abraham ST and Minkowski ST give only the conducting part of the total 'felt force', which can also be considered as the 'outside force' for point or microscopic objects such as electrons (for such objects, it is impossible to differentiate 'outside force' from 'interior force') |
| Einstein-Laub | $\overline{\overline{T}} = DE + BH - \frac{1}{2}(\mu_o H \cdot H + \varepsilon_0 E \cdot E)\overline{\overline{I}}$ <br><br> $f = (P \cdot \nabla)E + (M \cdot \nabla)H + \frac{\partial P}{\partial t} \times \mu_o H - \frac{\partial M}{\partial t} \times \varepsilon_0 E$ | $\frac{E \times H}{c^2}$ | The time averaged EL ST yields the 'felt force' or 'inside force' from fields just inside the object embedded in air or vacuum with no HQ*. Its force density distributes through the polarization $P$ and magnetization, $M$ just inside the object embedded in air or vacuum. This ST does not characterize the flow of momentum transported by the field from the background medium to an embedded scatterer. Hence it is not applicable to get the 'outside force'. |
| Chu | $\overline{\overline{T}} = \varepsilon_0 EE + \mu_0 HH - \frac{1}{2}(\mu_o H \cdot H + \varepsilon_0 E \cdot E)\overline{\overline{I}}$ <br><br> $f = -(\nabla \cdot P)E - (\nabla \cdot M)H + \frac{\partial P}{\partial t} \times \mu_0 H - \frac{\partial M}{\partial t} \times \varepsilon_0 E$ | $\frac{E \times H}{c^2}$ | The Chu ST and the force law show the 'existence domain' without HQ*. They determine just the interior kinetic part of the total force, along with the Abraham momentum density, by employing only the fields inside the object embedded in air. Only after manually adding the term HQ*, Chu force law obtains the total time-averaged 'felt force' from fields inside the object (and from the surface for HQ*). Like EL formulations, it is also not suitable to get a correct 'outside force' exerted on the embedded object from the background medium. |
| Lorentz/ Amperian/ Nelson | $\overline{\overline{T}} = \varepsilon_0 EE + \mu_0^{-1} BB - \frac{1}{2}\left(\mu_0^{-1} B^2 + \varepsilon_0 E^2\right)\overline{\overline{I}}$ <br><br> $f = -(\nabla \cdot P)E + \frac{\partial P}{\partial t} \times B + (\nabla \times M) \times B$ | $\varepsilon_0 E \times B$ | The Nelson ST and force law show the 'existence domain' without HQ1*. They are applicable to get the total 'felt force' from fields inside the object embedded in air or vacuum only after manually adding HQ* (surface fields for HQ1*) and HQ2*. Like EL and Chu formulations, it is not suitable to obtain the correct 'outside force' exerted on the embedded object from the background. |

[1]    HQ1* means hidden quantity in Chu's formulation (strictly the surface force term). Cf. Supplement S1 (a).
[2]    HQ2* means hidden quantity in the conventional Lorentz/Nelson force introduced by Shockley. Cf. Supplement S1 (b).
[3]    ** cf. G.B.Walker et al., Can.J.Phys **53**, 2577 (1975) and Rikken, G. L. J. A. et al., Phys. Rev. Lett. **108**, 230402 (2012).

### III. A unified theory: Defining three different systems to achieve a compact framework

Accoding to [24] only the total momentum, $p_{\text{Total}}$, resulting from adding that of the medium (kinetic, $p_{\text{kin}}^{\text{med}}$, or canonical, $p_{\text{Cono}}^{\text{med}}$) and the one corresponding to the field (Abraham's, $p_{\text{Abr}}$, or Minkowski's, $p_{\text{Mink}}$), is a conserved quantity:

$$p_{\text{Total}} = p_{\text{kin}}^{\text{med}} + p_{\text{Abr}}, \; p_{\text{Abr}} = \int G_{\text{Abr}} dv, \; G_{\text{Abr}} = \frac{E \times H}{c^2}; \tag{1}$$

$$p_{\text{Total}} = p_{\text{Cano.}}^{\text{med}} + p_{\text{Mink}}, \; p_{\text{Mink}} = \int G_{\text{Mink}} dv, \; G_{\text{Mink}} = D \times B. \tag{2}$$

$E$, $D$, $H$ and $B$ are the electric field, displacement, magnetic field and induction vectors, respectively [24]. $G_{\text{Abr}}$ and $G_{\text{Mink}}$ constitute the Abraham and Minkowski momentum densities of the field, respectively [4]. A problem is that Eqs (1) and (2) are generic equations; their appropriate applicability and their connections with the rigorous ST formulations are still unclear. To resolve this, and to unify all the previously mentioned STs, photon momenta and force laws of Table-1, as well as our generalization of the Einstein-Laub theory, we now define three different systems named as: (i) *'kinetic system'*, (ii) *'Minkowski system'* and (iii) the new one that we coin: *'generalized Einstein Laub system'*, (or *'GEL system'*). They are identified in Figs. 1 (a)-(c).

### III.A. Stress tensor and photon momenta of a 'kinetic system':

We first establish the ST and force law inside bodies embedded in air or vacuum. As seen in Fig. 1, we define *'kinetic systems'* as those in which Eq (1) applies describing the 'delivered momentum' inside a lossless or 'moderately absorbing' homogenous object embedded in air or vacuum [cf. Fig. 1 (a)]. By 'moderately absorbing' we mean the imaginary part ($\varepsilon_I$) of the permittivity $\varepsilon_s = \varepsilon_R + i\varepsilon_I$ of the scattering object reaches to a value after which the EL ST can not yield the total internal force. In our several numerical and full wave simulation results (i.e. Fig. 1s (a), (b) in supplement S0, Fig. 2s (a),(b) in supplement S4 (b), Fig. 3(a) in this main article and several other results [not shown in this work]), it is observed if $\varepsilon_I$ reaches the magnitude approximately ten times greater than the background permittivity ($\varepsilon_b$), EL ST (only when $\varepsilon_b = \varepsilon_0$) can yield the total internal force safely. We conclude that upto such an approximate limit of absorption, which can be defined as 'moderate absorption', Eq. (1) applies to the interior of a absorbing (and also lossless) object which therefore constitutes a *"transparent Einstein box"* [2,24]. This (from no loss to moderate absorption) can be considered as the region of interest for most of the realistic situations. As shown in Supplement S0, [Figs. 1s (a) and (b)], the interior of an absorbing dielectric or magneto-dielectric object supports only the EL ST up to a 'moderate loss' limit as discussed above. No other ST leads to the total internal force by employing the fields inside a scatterer. As a result, among all formulations [4] that convey the kinetic photon momentum $p_{Abr}$ [24], we support the EL formulations [22] excluding hidden momenta inside a moderately absorbing and homogenous dielectric or magneto-dielectric object embedded in air [1] [cf. Supplements S0 and S1 (a)-(c) for details]:

$$\left\langle \boldsymbol{F}_{Total}^{Kin.\ system} \right\rangle (in) = \oint \left\langle \bar{\bar{\boldsymbol{T}}}_{EL}^{in} \right\rangle \cdot d\boldsymbol{s} = \int \left\langle \boldsymbol{f}_{EL}^{in} \right\rangle dv. \qquad (3)$$

Here, $\bar{\bar{\boldsymbol{T}}}_{EL}^{in}$ is the EL ST and $\boldsymbol{f}_{EL}^{in}$ is the EL force density [4,21,22] (cf. Table-1), where 'in' stands for the field inside the body, and $\langle \cdot \rangle$ means time average addressing time-harmonic fields from now on. $\boldsymbol{P}_{Abr}$ (in) is the remaining electromagnetic part of the photon momentum, $\boldsymbol{p}_{electromag.}(in)$, after delivering the mechanical momentum of photon inside an object embedded in air. Though $\boldsymbol{P}_{Abr}$ (in) inside a scatterer may not be detectable in direct photon momentum based measurements such as a ray tracing method [7,33] or Doppler effect [4,35], $\boldsymbol{G}_{Abr}$ is emerged as the appropriate electromagnetic momentum density to obtain the correct distribution of the interior force. As a result, according to the linear momentum conservation equation: ($\int \boldsymbol{f}_{EL}^{in} dv + \frac{\partial}{\partial t} \int \boldsymbol{G}_{Abr}(in) dv = \oint \bar{\bar{\mathbf{T}}}_{EL}^{in} \cdot d\boldsymbol{s}$ and $\frac{\partial \boldsymbol{p}_{Kin}^{med}(in)}{\partial t} + \frac{\partial \boldsymbol{p}_{electromag.}(in)}{\partial t} = \frac{\partial \boldsymbol{p}_{Total}}{\partial t}$), Eq (1) should be considered as the law governing the conservation of total momentum for a 'kinetic system' so that the momentum delivery process inside any object embedded in air or vacuum, is described as:

$$\boldsymbol{p}_{Total} = \boldsymbol{p}_{Kin}^{med}(in) + \boldsymbol{p}_{Abr}(in). \qquad (4)$$

At this point we anticipate that later through Eq (21) we shall show that Eq (1) also governs the appropriate interior force and photon momentum of the embedding background if and only if the background is much larger than the embedded scatterer.

**III.B. Stress tensor and photon momenta of a 'Minkowski system':**

We now define '*Minkowski systems*' as those for which Eq (2) applies, describing the wave 'momentum transfer' [4] from the background to any generic embedded object, [cf. Fig.1 (b) and Fig.2].

**III.B. a. Minkowski stress tensor for the 'outside force'**

In contrast to some theoretical predictions [15,16,19,20,41,42], most experiments [3,7,13,14,32-35] have supported the Minkowski ST for the transfer of momentum from the background to an embedded object. Though Chu and EL theories in [19,20], and Nelson's formulation in [15,41,42], have been considered for the 'outside force'; those STs in the exterior region (i.e. outside the object, e.g. at $r = a^+$ for a sphere or cylinder) do not lead to a correct time-averaged force on the body embedded in the interface between two different backgrounds [33], nor even in a single medium [37, 38]. As a consequence, in general the 'outside force' can be appropriately calculated via the Minkowski ST:

$$\left\langle \boldsymbol{F}_{\text{Total}}^{\text{Mink. system}} \right\rangle (\text{out}) = \oint \left\langle \bar{\bar{\boldsymbol{T}}}_{\text{Mink}}^{\text{out}} \right\rangle \cdot d\boldsymbol{s}, \tag{5a}$$

$$\left\langle \bar{\bar{\boldsymbol{T}}}_{\text{Mink}}^{\text{out}} \right\rangle = \frac{1}{2}\text{Re}\left[ \boldsymbol{D}_{\text{out}} \boldsymbol{E}_{\text{out}}^* + \boldsymbol{B}_{\text{out}} \boldsymbol{H}_{\text{out}}^* - \frac{1}{2}\left( \boldsymbol{B}_{\text{out}} \cdot \boldsymbol{H}_{\text{out}}^* + \boldsymbol{D}_{\text{out}} \cdot \boldsymbol{E}_{\text{out}}^* \right)\bar{\bar{\boldsymbol{I}}} \right]. \tag{5b}$$

'out' stands for fields outside the object, [e.g., on $r=a^+$, if it is a sphere or cylinder of radius *a*, cf. Fig. 1(b)] and $\bar{\bar{\boldsymbol{I}}}$ is the unity tensor. The electromagnetic vectors in (5b) correspond to the total field, namely, incident plus field scattered by the body. The force in Eq (5a) is defined as 'outside force' just for convenience, as $\bar{\bar{\boldsymbol{T}}}_{\text{Mink}}^{\text{out}}$ characterizes only the momentum transfer rather than the real force 'felt' by the embedded object. In the next sections we validate previous experimental observations [3,7,13,14, 33-35] and theoretical predictions [36-40] of Minkowski's theory for the 'outside force' by employing only the interior field of an embedded object with our generalized EL equations.

### III.B.b. Transfer of the Minkowski photon momentum from the background

Most major experimental works regarding photon momentum are based on changes in space rather than in time. This is one of the reasons why the transfer of Abraham photon momentum $p_{Abr}(\text{out})$ from the background to an embedded object is not clearly detectable [43], (cf. also our note [44]). As a result, based on the previous observations in favor of $p_{Mink}$ (out) [7, 24, 33, 36] and according to several numerical results shown below in this paper, (specially in section IV), supporting Minkowski's ST, $p_{Mink}$ (out) should be considered the measurable transferred 'total' photon momentum, or wave momentum [4], from the background to the object. The term 'total' means that $p_{Mink}$ (out), which arises on the boundaries during the momentum transfer [40], is the sum of the exterior electromagnetic momentum $p_{Abr}(\text{out})$ plus a mechanical momentum: $p_{med}(\text{out}) = (n_b - n_b^{-1})\hbar\omega/c$ [33] carried by the field [4, 33].

But, what is the connection of Eq. (2) with $p_{Mink}$ (out)?. According to the quantum nature of light [45] $p_{Mink}(=\hbar k)$ is associated to the canonical momentum of photons [2, 24], ($\hbar$ being the reduced Plank constant and $k$ representing the wavevector of light inside the medium). Though no difference is observed between the time-averaged external Abraham and Minkowski STs and the forces (cf. Table-1), the important unnoticed physical difference between $G_{Abr}(\text{out})$ and $G_{Mink}(\text{out})$:

$$[\frac{\partial}{\partial t}\int G_{Mink}(\text{out})dv = (\frac{\partial}{\partial t}\int G_{Abr}(\text{out})dv + \int f_A(\text{out})dv) = (\frac{\partial p_{Abr}(\text{out})}{\partial t} + \frac{\partial p_{med}(\text{out})}{\partial t})]$$

can be reinterpreted according to our notes [46] and [47]. Here the extra Abraham term $f_A(\text{out})$ (cf. Table-1 and note [47]) should be considered as the transferred mechanical part (transported by the field) involved in $G_{Mink}(\text{out})$. This interpretation also supports the canonical momentum transfer of photons according to notes [46] and [47]. As a result, in accordance with Eq (2) the equation governing the total momentum and describing the transfer of wave momentum from the background in a 'Minkowski system' should be written as:

$$p_{Total} = p_{Cano.}^{med}(\text{out}) + p_{Mink}(\text{out}). \tag{6}$$

**III. C. Stress tensor, force and photon momentum of a 'generalized Einstein Laub (GEL) system'**

The Minkowskian formulation cannot describe how the force is distributed (i.e., how it is 'felt') inside an arbitrary object immersed in a generic background, [cf. Fig. 1 (b) and (c) and Fig.2]. Even the EL formulation (cf. Table-1) that supports the $p_{Abr}(in)$ with fields inside the object, (e.g., at $r = a^-$ for a sphere or cylinder) cannot deal with complex configurations [3,7,13,14,19,20,26,29,31,33,36-40]. Ref. [24] concludes: "*By demonstrating the need for two "correct" momenta and associating these, unambiguously, with the Abraham and Minkowski forms, we may hope that we have also removed the need for further rival forms.*" But according to our analysis, though the measurent of the transferred photon momentum can be considered to yield the Minkowski one for different cases, the measurent of the total internal force (or force distribution) in different systems may lead to different forms of non-mechanical internal momenta other than the one of Abraham. This is the reason why we have introduced the name 'Generalized Einstein-Laub (GEL) system' as that in which our GEL equations established below should lead to the appropriate total force and momentum inside the object. It is even useful to distinguish what we call 'GEL systems of first kind' from those that we name 'GEL systems of second kind'. In the former one deals with the internal dynamics of a lossless or moderately absorbing object (see above) embedded in another material background, (from now on when we refer to an embedded object, unless explicitly stated we shall mean such a lossless or moderately absorbing body). By contrast, in the latter one addresses the internal dynamics of an extremenly absorbing object embedded in air or vacuum. [Here by 'extremely absorbing' we mean an imaginary part $(\varepsilon_I)$ of the scatterer permittivity $\varepsilon_s = \varepsilon_R + i\varepsilon_I$, which is much higher (i.e. $\varepsilon_I \gg \text{background permittivity}, \varepsilon_b$) than the previously described vale of $\varepsilon_I$ for the case of 'moderate absorption'.We discuss such an extreme situation in Section VI and Supplement S4 (d)]. Finally, we define as 'GEL systems of third kind' those in which we consider the internal dynamics of a chiral object embedded in another material background. We shall show our previously defined 'kinetic system' and the wel-known Einstein-Laub equations are only special cases of a 'GEL system' and our GEL equations respectively.

**III.C. a. Physical quantities for a 'GEL system': The generalized Einstein-Laub equations**

In supplements S2(a) and (b) we show that the Einstein-Balaz's type thought experiment may not lead to the appropriate internal photon momentum of an embedded object (i.e. this is a 'GEL system of first kind'). On the other hand, the arguments in favor of non-relativistic Doppler-shift effects [24, 48] always lead to the transfer of photon momentum from the background to an embedded object rather than to the internal photon momentum of the object. Hence, instead of following those thought experiments, the internal photon momentum of the 'GEL system' should be determined by a more rigorous approach based on the linear momentum conservation equation: $\nabla \cdot \bar{\bar{T}} = f + \frac{\partial}{\partial t} G$.

When the embedding medium is magneto-dielectric instead of air, or the internal loss of the object crosses the limit of 'moderate absorption', the conventional vectors $P$ and $M$ inside the object should be replaced by what we shall from now on call 'effective polarization' $P_{\text{Eff}}$ and 'effective magnetization' $M_{\text{Eff}}$, respectively. We will show several instances where the conservation of linear momentum agrees with the use of $P_{\text{Eff}}$ and $M_{\text{eff}}$, rather than of the conventional ones. The justification of these two quantities lies on the fact that in such cases, though the conventional induced polarization and magnetization currents of the EL equations take place inside the object, they only partially contribute to the time-averaged optical force. In other words, the usual $P$ and $M$ introduced in the EL equations are unable to handle complex situations [3,7, 10, 13, 14, 20,26,29,31,33,36-40]. However, as shown next the modification here introduced is general, and satisfactorily deals with all numerical experiments that we have set on trial. Hence for the 'GEL system', we establish the following 'generalized constitutive relations' inside a generic object immersed in an arbitrary medium:

$$D_{\text{in}} = D_{\text{Eff}} = \varepsilon_{\text{Eff}} E_{\text{in}} + P_{\text{Eff}}; \quad P_{\text{Eff}} = (\varepsilon_S - \varepsilon_{\text{Eff}}) E_{\text{in}}, \tag{7a}$$

$$B_{\text{in}} = B_{\text{Eff}} = \mu_{\text{Eff}} H_{\text{in}} + M_{\text{Eff}}; \quad M_{\text{Eff}} = (\mu_S - \mu_{\text{Eff}}) H_{\text{in}}. \tag{7b}$$

*Eqs. 7(a) and (7b) constitute the first key proposal of this work.*

The constitutive parameters $\mu_{\text{Eff}}$ and $\varepsilon_{\text{Eff}}$ depend on the type of embedding medium and also on the object optical properties characterized by its permittivity and permeability, $\varepsilon_S$ and $\mu_S$, [see examples in Supplement S4 (a)-(d)].

Prior to establishing our unified formulation, we put forward a postulate formulated in note [49], based on which we establish the following generalization of the Einstein-Laub (GEL) stress tensor in terms of the vectors $\boldsymbol{P}_{\text{Eff}}$ and $\boldsymbol{M}_{\text{Eff}}$ as follows:

$$\bar{\bar{\boldsymbol{T}}}^{in}_{\text{GEL}} = \bar{\bar{\boldsymbol{T}}}^{in}_{\text{Mink}} + \frac{1}{2}\left(\boldsymbol{M}_{\text{Eff}} \cdot \boldsymbol{H}_{in} + \boldsymbol{P}_{\text{Eff}} \cdot \boldsymbol{E}_{in}\right)\bar{\bar{\boldsymbol{I}}}, \tag{8a}$$

Here the internal Minkowski ST is $\bar{\bar{\boldsymbol{T}}}^{in}_{\text{Mink}} = \boldsymbol{D}_{in}\boldsymbol{E}_{in} + \boldsymbol{B}_{in}\boldsymbol{H}_{in} - \frac{1}{2}\left(\boldsymbol{B}_{in} \cdot \boldsymbol{H}_{in} + \boldsymbol{D}_{in} \cdot \boldsymbol{E}_{in}\right)\bar{\bar{\boldsymbol{I}}}$. Note that $\bar{\bar{\boldsymbol{T}}}^{in}_{\text{Mink}}$ in Eq (8a) has to be expressed in terms of the interior fields of the scatterer, whereas $\bar{\bar{\boldsymbol{T}}}^{out}_{\text{Mink}}$ in Eq (5a) is given by the exterior background fields of the body. From Eq (7a) and (7b) and postulate [49], Eq (8a) can be written for any generic object as:

$$\bar{\bar{\boldsymbol{T}}}_{\text{GEL}} = \boldsymbol{D}_{in}\boldsymbol{E}_{in} + \boldsymbol{B}_{in}\boldsymbol{H}_{in} - \frac{1}{2}\left(\mu_{\text{Eff}}\boldsymbol{H}_{in} \cdot \boldsymbol{H}_{in} + \varepsilon_{\text{Eff}}\boldsymbol{E}_{in} \cdot \boldsymbol{E}_{in}\right)\bar{\bar{\boldsymbol{I}}} \tag{8b}$$

*Eqs.(8a) and (8b) are the second key proposal of this paper.*

Notice that Eq (8a), or (8b), is the root [49] to obtain the time averaged force for the different 'GEL systems' of 'first', 'second', or 'third kind', especially where the non-mechanical photon momentum cannot be considered as that of Abraham. In such situations, one just needs to appropriately introduce the corresponding $\boldsymbol{P}_{\text{Eff}}$ and $\boldsymbol{M}_{\text{Eff}}$, or equivalently $\mu_{\text{Eff}}$ and $\varepsilon_{\text{Eff}}$, to correctly formulate those different cases. In order to satisfy the linear momentum conservation, a non-mechanical momentum density can be introduced. This should yield a time-averaged force density that fulfills two simple test equations for different cases:

$$\langle \boldsymbol{F}_{\text{Total}} \rangle (\text{in}) = \oint \langle \bar{\bar{\boldsymbol{T}}}^{in} \rangle \cdot d\boldsymbol{s} = \int \langle \boldsymbol{f}^{in} \rangle dv \tag{9a}$$

$$\langle \boldsymbol{F}_{\text{Total}} \rangle (\text{out}) \approx \langle \boldsymbol{F}_{\text{Total}} \rangle (\text{in}). \tag{9b}$$

*Eq (9b) is the third key proposal of this work,* constituting the test for several different set-ups throughout this paper. This will verify the 'existence domain' of the different STs, force laws and photon momenta.

We show in Supplements S3 (a) and (b) that if one considers the most appealing non-mechanical momentum density: $G_{GEL}^{Transition}(in) = [D_{in} - P_{Eff}] \times [B_{in} - M_{Eff}] = \varepsilon_{Eff}\mu_{Eff}(E_{in} \times H_{in})$, (that turns into $G_{Abr}(in)$ if the background of a scatterer is air or vacuum), it leads to a force density: $f_{GEL}^{Transition} = (P_{Eff} \cdot \nabla)E_{in} + (M_{Eff} \cdot \nabla)H_{in} + \frac{\partial P_{Eff}}{\partial t} \times \mu_{Eff}H_{in} - \frac{\partial M_{Eff}}{\partial t} \times \varepsilon_{Eff}E_{in}$. However, by testing different cases we have found that this force does not fulfill the aformentioned two test equations (9a) and (9b) for the time-averaged force. Now, we consider that the ST in Eq (8b) remains unchanged at any instant according to the equation: $\nabla \cdot \bar{\bar{T}} = f + \frac{\partial}{\partial t}G$. By substracting the $\partial(P_{Eff} \times M_{Eff})/\partial t$ term from $\partial G_{GEL}^{Transition}(in)/\partial t$ and then adding it with $f_{GEL}^{Transition}$, while maintaining $\nabla \cdot \bar{\bar{T}}$ fixed, we finally arrive to our generalized Einstein-Laub time-averaged force which complies with the two test equations (9a) and (9b) for different cases. Hence to determine the correct force felt by any generic scatterer, one should employ the law:

$$\langle f_{GEL}^{in} \rangle = \frac{1}{2}\text{Re}\left[(P_{Eff} \cdot \nabla)E_{in}^* + (M_{Eff} \cdot \nabla)H_{in}^* - i\omega(P_{Eff} \times B_{in}^*) + i\omega(M_{Eff} \times D_{in}^*)\right].$$
(10a)

$\langle f_{GEL}^{in} \rangle$ is the time-averaged generalized Einstein-Laub force density. *Eq (10a) is the fourth key proposal of this paper.* By applying Eqs (7a) and (7b), we get from Eq (10a):

$$\langle f_{GEL} \rangle = \frac{1}{2}\text{Re}[((\varepsilon_S - \varepsilon_{Eff})E_{in} \cdot \nabla)E_{in}^* + ((\mu_S - \mu_{Eff})H_{in} \cdot \nabla)H_{in}^*$$
$$- i\omega((\varepsilon_S - \varepsilon_{Eff})E_{in} \times B_{in}^*) + i\omega((\mu_S - \mu_{Eff})H_{in} \times D_{in}^*)].$$
(10b)

where $B_{in}$ and $D_{in}$ should be written as $\mu_s H_{in}$ and $\varepsilon_s E_{in}$, respectively.

Now, the non-mechanical momentum density in the 'GEL system' that supports Eq (10a) should be written [subtracting the $\partial(P_{Eff} \times M_{Eff})/\partial t$ term from $G_{GEL}^{Transition}(in)$] as:

$$G_{\text{GEL}}(\text{in}) = [[(D_{\text{in}} - P_{\text{Eff}}) \times (B_{\text{in}} - M_{\text{Eff}})] - (P_{\text{Eff}} \times M_{\text{Eff}})] \tag{11a}$$

*Eq (11a) is the fifth key proposal of this work*. Based on Eqs (7a) and (7b), we can write Eq (11a) as:

$$G_{\text{GEL}}(\text{in}) = (\varepsilon_{\text{Eff}}\mu_s + \varepsilon_s\mu_{\text{Eff}} - \varepsilon_s\mu_s)(E_{\text{in}} \times H_{\text{in}}) \tag{11b}$$

Though in Supplements S2 (a) and (b) it is shown that the internal photon momentum of an embedded Einstein-Balazs box manifests as the one of Abraham; any force or ST related to $p_{\text{Abr}}(\text{in})$, does not fulfill the test Eqs (9a) and (9b). An explanation for this is that $p_{\text{Abr}}(\text{in})$ is the pure electromagnetic part of the total non-mechanical photon momentum $p_{\text{Non-Mech.}}(\text{in})$, which remains a fixed mathematical expression inside any generic scatterer. But in general situations, Eq (11b) should express the generalized 'total' non-mechanical momentum density and, considering $G_{\text{Abr}}$ as the electromagnetic momentum density $G_{\text{Electromag}}$, it can be splitted into two terms as:

$$G_{\text{GEL}}(\text{in}) = \left[ G_{\text{Material}} + G_{\text{Electromag}} \right] = [c^2(\varepsilon_{\text{Eff}}\mu_s + \varepsilon_s\mu_{\text{Eff}} - \varepsilon_s\mu_s) - 1](E_{\text{in}} \times H_{\text{in}})/c^2 + (E_{\text{in}} \times H_{\text{in}})/c^2 \tag{12a}$$

Therefore a 'material induced momentum' $p_{\text{Material}}(\text{in}) = \int G_{\text{Material}} dv$ is always added to the pure electromagnetic photon momentum inside a generic scatterer; where

$G_{\text{Material}} = [[c^2(\varepsilon_{\text{Eff}}\mu_s + \varepsilon_s\mu_{\text{Eff}} - \varepsilon_s\mu_s) - 1](E_{\text{in}} \times H_{\text{in}})/c^2]$. Hence Eq (12a) can be expressed as:

$$p_{\text{Non-Mech.}}(\text{in}) = p_{\text{Material}}(\text{in}) + p_{\text{Abr}}(\text{in}) \tag{12b}$$

*Eq (12a), or (12b), is the sixth key proposal of this work*. From them it is clear that $p_{\text{Material}}(\text{in})$ is a function of $\varepsilon_{\text{Eff}}$ and $\mu_{\text{Eff}}$. In this paper we show different GEL systems for which $\varepsilon_{\text{Eff}}$ and $\mu_{\text{Eff}}$ take different appropriate values to satisy Eq (9b). As a result, considering this nature of $p_{\text{Material}}(\text{in})$, [and hence of $p_{\text{Non-Mech.}}(\text{in}) = \int G_{\text{GEL}}(\text{in}) dv$ in Eq (12b)], the generalized total momentum conservation equation for a 'GEL system' should be expressed as:

$$p_{\text{Total}} = p_{\text{Mechanical}}^{\text{med}}(\text{in}) + p_{\text{Non-Mech.}}(\text{in}). \tag{13}$$

*Eq(13) is the seventh key proposal of this work.* There $p_{\text{Mechanical}}^{\text{med}}(\text{in})$ and $p_{\text{Non-Mech.}}(\text{in})$ are respectively the mechanical momentum of a generic scatterer and the non-mechanical photon momentum inside this body at any instant. We shall show by means of several illustrations that Eq. (4) should be considered a special case of Eq (13) when the value $\varepsilon_l$ of the object embedded in air or vacuum does not cross beond the limit of moderate loss. However, the question is how Eq (8b) and (10b) satisfy the test Eq (9b). This is answered in the following sections and in supplement S4 (a)-(d) considering several complex situations.

### III.C. b. GEL system of first kind: Introducing the first kind generalized Einstein-Laub equations

For the 'GEL system of first kind' defined previously, [cf. Fig. 1 (c)], coresponding to bodies with no, or moderate, absorption, embedded in an arbitrary medium, we identify the constitutive parameters of Eqs. (7a) and (7b) as those of the immediate background region: $\varepsilon_{\text{Eff}} = \varepsilon_b$ and $\mu_{\text{Eff}} = \mu_b$, (i.e that region sharing its interfaces with the embedded body). For these GEL systems of first kind we find useful to rename the generalized Einstein-Laub equations as 'first kind GEL (or 'GEL1') equations'. Hence from Eq (8b), the first kind GEL (or 'GEL1') ST can be written as:

$$\bar{\bar{T}}_{\text{GEL1}} = D_{\text{in}} E_{\text{in}} + B_{\text{in}} H_{\text{in}} - \frac{1}{2}\left(\mu_{b(j)} H_{\text{in}} \cdot H_{\text{in}} + \varepsilon_{b(j)} E_{\text{in}} \cdot E_{\text{in}}\right)\bar{\bar{I}}. \tag{14}$$

where *j=1,2,3,...., N* represents the number of background regions sharing interface with the object, [cf. Figs. 2(a) and 2(b)]. *Eq (14) is the eighth key proposal of this work.*

When the background is vacuum or air, our first kind GEL stress tensor [Eq (14)] turns into the Einstein-Laub ST. As a result, we conclude that the Einstein-Laub ST is only a special case of the GEL1 ST put forward in this paper.

On the other hand, to fulfill the linear momentum continuity equation ( $f = \nabla \cdot \bar{\bar{T}} - \frac{\partial}{\partial t} G$ ) inside an embedded object, Eq. (12a) leads to the first kind GEL non-mechanical momentum density:

$$G_{\text{GEL1}} = \left[G_{\text{Material}} + G_{\text{Electromag}}\right] = [c^2(\varepsilon_{b(j)}\mu_s + \varepsilon_s\mu_{b(j)} - \varepsilon_s\mu_s) - 1](E_{\text{in}} \times H_{\text{in}})/c^2 + (E_{\text{in}} \times H_{\text{in}})/c^2 \tag{15}$$

like in Eq (14), *j=1,2,3,...., N* representing the number of background regions sharing interface with the object. According to our discussion in Sub-Section III.C.a, the equation governing the conservation of total momentum in a 'GEL system of first kind' should be written as:

$$\boldsymbol{p}_{\text{Total}} = \boldsymbol{p}_{\text{Mechanical}}^{\text{med}}(\text{in}) + \boldsymbol{p}_{\text{Non-Mech.}}(\text{in}). \tag{16}$$

Here $\boldsymbol{p}_{\text{Mechanical}}^{\text{med}}(\text{in})$ and $\boldsymbol{p}_{\text{Non-Mech.}}(\text{in}) = \int \boldsymbol{G}_{\text{GEL}}^{\text{1st}} dv$ are respectively the mechanical and the non-mechanical momentum at any instant of the embedded scatterer, (i.e. the embedded Einstein-Balazs box). Eq. (4) reveals as a special case of Eq (16). However, according to Eq (10b) the interior time-averaged force on an embedded scatterer should be written as:

$$\langle \boldsymbol{f}_{\text{GEL1}} \rangle = \frac{1}{2} \text{Re}[((\varepsilon_S - \varepsilon_{b(j)})\boldsymbol{E}_{\text{in}} \cdot \nabla)\boldsymbol{E}_{\text{in}}^* + ((\mu_S - \mu_{b(j)})\boldsymbol{H}_{\text{in}} \cdot \nabla)\boldsymbol{H}_{\text{in}}^* \\ - i\omega((\varepsilon_S - \varepsilon_{b(j)})\boldsymbol{E}_{\text{in}} \times \boldsymbol{B}_{\text{in}}^*) + i\omega((\mu_S - \mu_{b(j)})\boldsymbol{H}_{\text{in}} \times \boldsymbol{D}_{\text{in}}^*)]. \tag{17}$$

*Eq (17) is the ninth key proposal of this paper.* Note that when $\varepsilon_b = \varepsilon_0$ and $\mu_b = \mu_0$, the time-varying version of Eq (17) does not turn into the Einstein-Laub force due to the presence of the $\frac{\partial \boldsymbol{P}}{\partial t} \times \boldsymbol{B}_{\text{in}} - \frac{\partial \boldsymbol{M}}{\partial t} \times \boldsymbol{D}_{\text{in}}$ term, that appears instead of $\frac{\partial \boldsymbol{P}}{\partial t} \times \mu_o \boldsymbol{H} - \frac{\partial \boldsymbol{M}}{\partial t} \times \varepsilon_0 \boldsymbol{E}$, (cf.Table-1). However, it is interesting that Eq. (17) also leads to the correct time-averaged internal force on an object embedded in air or vacuum with moderate absorption, (i.e. a 'kinetic system', discussed in Sub-Section III.A). Thus the time-averaged force felt by an embedded scatterer should be expressed in general as:

$$\langle \boldsymbol{F}_{\text{Total}}^{\text{GEL1 system}} \rangle(\text{in}) = \oint \langle \overline{\overline{\boldsymbol{T}}}_{\text{GEL1}}^{\text{in}} \rangle \cdot d\boldsymbol{s} = \int \langle \boldsymbol{f}_{\text{GEL1}}^{\text{in}} \rangle dv. \tag{18}$$

Thus the previously defined 'kinetic system' reveals to be a special case of a 'GEL system of first kind'.

**IV. Additional features of 'GEL systems of first kind': The internal dynamics of an object embedded in various types of background media**

**IV.A. Homogenous unbounded background and the experiment on trapping of biological cells**

Our GEL1 ST, Eq (14), yields the total force from fields inside an arbitrary object, [i.e. either absorbing, gain, dielectric, or magneto-dielectric, with e.g. slab, sphere, or cylinder shape; cf. Fig. 1 (c)] embedded in a generalized magneto-dielectric medium. For example, on an absorbing magneto-dielectric 3D scatterer embedded in a homogeneous magneto-dielectric background, it is illustrated in Fig. 3a that the 'outside force' given by Minkowski's ST is in full agreement with the interior force (which is the 'force felt' by the scatterer) given by GEL1 ST up to the limit of moderate absorption, (i.e. $(\varepsilon_I / \varepsilon_b) < 10$). In addition, several other results from Mie calculations and full wave simulations support such agreements, which are illustrated in Supplements S4 (a)-(d), as well as in the next sections. In general, for arbitrary scatterers with moderate, low, or no absorption at all, from Eq (5a) and Eq (18) we have:

$$\left\langle \boldsymbol{F}_{\text{Total}}^{\text{Mink. system}} \right\rangle (\text{out}) \approx \left\langle \boldsymbol{F}_{\text{Total}}^{\text{GEL1 system}} \right\rangle (\text{in}). \tag{19}$$

*Eq (19) is the tenth key proposal of this work,* [cf. also Eq (9b)].

Especially, we remark the importance of the slab example illustrated in Supplement S4 (a). Let us consider a lossless magneto-dielectric slab, embedded in another lossless unbounded magneto-dielectric medium, illuminated at normal incidence by a linearly polarized plane wave propagating along the z-direction with electric vector: $E_x = E_0 e^{i(kz-\omega t)}$. The 'outside force' can be calculated considering the transfer of $\boldsymbol{p}_{\text{Mink}}(\text{out}) = \hbar \boldsymbol{k}$ from the background to the embedded slab (cf. Eq (60) in [50]). At the same time, the interior force on the embedded slab is calculated by our first kind GEL ST [cf. Eq. (14)] and the force density formula of Eq. (17). All of them are in full agreement with each other:

$$\left[ N_i \hbar k_i + N_r \hbar k_r - N_t \hbar k_t \right]_{\text{From Background}}^{\text{Transfer Process}} = \frac{1}{2} \frac{E_0^2}{\mu_s \left( \frac{\mu_b \varepsilon_S}{\mu_S \varepsilon_b} - 1 \right)} \left[ (\varepsilon_s - \varepsilon_b) \mu_s - (\mu_s - \mu_b) \varepsilon_s \right] \left\{ 1 + |R|^2 + |T|^2 \right\} \tag{20a}$$

In Eq.(20a), '*i*', '*r*' and'*t*' mean incident, reflected and transmitted, respectively. Thus the left side of (20a) describes the transfer of momentum to the embedded slab employing the incident, reflected, and transmitted $\boldsymbol{p}_{\text{Mink}}(\text{out}) = \hbar \boldsymbol{k}$, respectively. In contrast, the right side of (20a) represents the total internal GEL force felt by the slab, obtained by employing only its internal field. The quantity

$N = \langle S \rangle / (\hbar \omega)$ denotes the photon flux and $\langle S \rangle$ is the time-averaged Poynting vector. $R$ and $T$ stand for the reflection and transmission coefficients of the slab. In consequence, with the aid of Eq (20a) one can write Eq (19) for a slab as:

$$\langle \boldsymbol{F}_A^{\text{Mink. system}} \rangle (\text{out}) = \langle \boldsymbol{F}_A^{\text{GEL1 system}} \rangle (\text{in}), \tag{20b}$$

where $\boldsymbol{F}_A$ is the force per unit area on the slab. Our calculations verify that the internal time-averaged first kind GEL force, Eq (17), acts on the magneto-dielectric particle by interaction between $\boldsymbol{J}^{\text{Electric}}$ ($= \partial \boldsymbol{P}_{\text{Eff}} / \partial t$) and $\boldsymbol{B}$ [23,51] (instead of $\mu_0 \boldsymbol{H}_{\text{in}}$ or $\mu_b \boldsymbol{H}_{\text{in}}$); and between $\boldsymbol{J}^{\text{Magnetic}}$ ($= \partial \boldsymbol{M}_{\text{Eff}} / \partial t$) and $\boldsymbol{D}$. This resolves long lasting [4] debates regarding the EL force as pointed in [23] as well as some confusions between the time-varying and the time-averaged forces reported in [51].

For the biological cell trapping experiment [26], Chu [27], Einstein-Laub [4] and Minkowski's forces [4] lead to the same magnitude of the force considering a simplified model based on a slab embedded in air [4, 27]. However, we must state that even without considering such simplified models, and even without employing any hidden quantity, Eq (20a) matches the exact results reported in [26] on considering the actual non-vacuum background (cf. Eqs (1a) and (1b) in [26] which involve the transfer of momentum $p_{\text{Mink}}$ from the background to the object). Modelling the cell as a dielectric slab embedded in a dielectric, as done in the actual work [26], the interior force can be correctly calculated from the first kind GEL ST, Eq (14), but not via Chu or any other ST. Even if the cell is replaced by an arbitrarily shaped magneto-dielectric scatterer embedded in another magneto-dielectric background, our conclusions based on Eq (19) remain valid, as shown in Fig. 3(a), and also in Supplements S4 (a)-(d) and in the next examples.

**IV. B. Heterogeneous unbounded background: Force dynamics in an embedded object**

Fig. 3(b) illustrates that the first kind GEL ST yields the correct interior force on an embedded scatterer even if the background medium is heterogeneous, [see also Fig. 2(a)]. As seen, Eq. (19) remains valid not only for the simple case of a dielectric object submerged in a bi-background: air-water [7, 33], but also for an infinite magneto-dielectric circular cylinder embedded in a heterogeneous medium of four different magneto-dielectric layers, as illustarted in Fig. 3(b). There the object is a 2D magneto-dielectric cylinder, with radius $a$=2000 nm. 25% of the particle is immersed in each of the four magneto-dielectric backgrounds. Our first kind GEL ST is employed in regions 1, 2, 3, and 4, inside the embedded object by using only interior fields, [cf. Fig. 3s in Supplement S4(c)], but with four different permittivities and permeabilities of the backgrounds employed in this first kind GEL ST, Eq (14). These complex situations are correctly tackled by our 2D full wave simulation. More details on the connection between the first kind GEL ST and our generalized Einstein-Laub theory with a heterogeneous background are highlighted in Supplement S4 (c).

### IV. C.  A first example of bounded background: Force dynamics in an embedded core

Our formulation is further verified by obtaining the force on a dielectric sphere, with core radius $a$, coated by a dielectric shell (core-shell total radius: $b$) embedded in air. For $a$=50 nm and $b$=60 nm, [cf. Fig. 2(b) and the inset in Fig. 4(a) illustrating the bounded background]. The time-averaged 'outside force' on the core, $\langle F_{\text{out}}^{\text{Core}} \rangle$, which is calculated by employing Minkowski ST, Eq (5), on $r=a^+$ with fields in the shell, conveys the transfer of $p_{\text{Mink}}$ (out) from the shell to the embedded core. This corresponds to a 'Minkowski system' defined in section III.B.a. Fig. 4(a) illustrates that the 'outside force' is in full agreement with our first kind GEL ST based time-averaged force, Eq (14), here denoted as $\langle F_{\text{in}}^{\text{Core}} \rangle$, using the *fields inside the core* [at $r = a^-$]. Eq (19) remains also valid for this configuration, [cf. Fig.4 (a)]. In consequence, the equation governing the conservation of total momentum inside the embedded core should be written from Eq (16) as [cf. also Fig.1 (c)]: $p_{\text{Total}} = p_{\text{Mechanical}}^{\text{core}}(\text{in}) + p_{\text{Non-Mech.}}(\text{in})$. Where $p_{\text{Mechanical}}^{\text{core}}(\text{in})$ is the mechanical momentum of the core and $p_{\text{Non-Mech.}}(\text{in}) = \int G_{\text{REL}}(\text{in}) dv$.

## V. Force dynamics in the embedding background

### V. A. A second example of bounded background

So far we have only considered the issues concerning the scattering object. Now we propose a thought experiment to gain insight on the appropriate force distribution and photon momentum in the embedding background. Let us consider another core-shell example with $b \gg a$, [cf. Fig. 4(b) with $b=8a$], where the whole core-shell object embedded in air can be considered almost homogeneous, (i.e., this is a kinetic system). The time-averaged outer force $\langle F_{\text{out}}^{\text{Shell}} \rangle$ on this entire body, obtained from the air fields on $r = b^+$, by the conventional Maxwell ST, (which is the Minkowski ST), is in full agreement [cf. Fig.4(b)] with the time-averaged force $\langle F_{\text{in}}^{\text{Shell}} \rangle$ derived from the EL ST on employing the interior fields of the whole object on $r = b^-$. Notice that the EL ST supports $p_{\text{Abr}}(\text{in})$ in the entire body, being mainly constituted by the large shell, and which can be considered as the background of the tiny core. In consequence, if and only if the background is much larger than the embedded scatterer, the equation goberning the conservation of total momentum in that background can be written in accordance with Eq (1) as:

$$p_{\text{Total}} = p_{\text{kin}}^{\text{Background}}(\text{in}) + p_{\text{Abr}}(\text{in}). \qquad (21)$$

$p_{\text{kin}}^{\text{Background}}(\text{in})$ is the kinetic momentum of such a large background. Of course Eq. (21) is valid only when the embedded scatterer (i.e. the tiny core) is small enough so that its presence can be neglected, [cf. Fig.1 (a)]. Note that in the previous core-shell example of Section IV. C the core-shell object is overall inhomogeneous, ($a=50$ nm; $b=60$ nm). Hence, the interior force on the entire object cannot be calculated by the EL ST, nor the associated photon momentum in the shell can be defined as $p_{\text{Abr}}(\text{in})$.

### V. B. Important conclusions regarding the two photon momenta in the same background

The resolution of the dilemma of two photon momenta in the same background can be well understood by recognizing the completely different roles of $\boldsymbol{p}_{\text{Abr}}(\text{in})$ and $\boldsymbol{p}_{\text{Mink}}(\text{out})$ in Eqs. (21) and (6), respectively. 'The dilemma should not be: which photon momentum is correct and which one is incorrect? Rather it should be: in a specific measurement, which photon momentum should emerge and which should not?'. Then our resolution of the dilemma is:

*"If a measurement is done only to determine the transfer of optical momentum into an embedded body, Minkowski photon momentum emerges at the interfaces or boundaries of the body. In contrast, if a measurement is done to detemine the internal force (and the force distribution) in the embedding background, then the electromagnetic photon momentum in the large background medium emerges as that of Abraham"*

In the example of Section V.A, the large shell can be considered as the unbounded liquid background of the experiment reported in [34], while the embedded core in the examples of Sections IV.C and V.A plays a role similar to the embedded mirror of [34]. The *measured* force felt by the large shell, and the photon momentum there, (i.e. like in the water background of [34], cf. also Eq (21) in [20]), reveals to be the internal EL force and $\boldsymbol{p}_{\text{Abr}}(\text{in})$, respectively. In contrast, the *measured* 'outside force' on the core, (like the transfer of momentum to the embedded mirror in [4, 34]), as well as the photon momentum there, manifests as that from the external Minkowski force [3], and $\boldsymbol{p}_{\text{Mink}}(\text{out})$, respectively, [35]. I.e. according to the left side of our Eq (20a) if the permittivity of the embedding medium is increased by $\sqrt{n_b}$ and the incidence angle of the beam is $\theta_i$, the force on such a perfect reflector in [34], modelled as a slab [3,20,35], (i.e with $|T|^2 = 0$ and $|R|^2 = 1$), increases from $\varepsilon_0 E_0^2 \cos^2\theta_i$ to $\sqrt{n_b}\varepsilon_0 E_0^2 \cos^2\theta_i = (2n_b/c)\langle S\rangle \cos^2\theta_i$ supporting the transfer of momentum $\boldsymbol{p}_{\text{Mink}}(\text{out})$, as observed in [34]. Three other major radiation pressure experiments [7, 52, 53] also support our above resolution of the dilemma. We discuss them in our footnotes [54] and [55].

**VI. GEL system of second kind: Internal dynamics in a highly absorbing object embedded in air**

We return to the case of a highly absorbing scatterer, for which the EL ST fails to yield the total interior force even if the body is embedded in air, [cf. our discussion on 'kinetic systems' in section III.A]. We show that then the internal dynamics of such objects is different to that of a scatterer embedded in a material background, [cf. our discussion in the beginning of section III.C]. For an arbitrary dielectric or magneto-dielectric object embedded in air, we can write:

$$\langle \bar{\bar{T}}_{\text{GEL}} (\text{in}) \rangle = \langle \bar{\bar{T}}_{\text{EL}} (\text{in}) \rangle, (\varepsilon_{\text{Eff}}, \mu_{\text{Eff}}) = (\varepsilon_0, \mu_0), \text{when moderately absorbing};  \quad (22a)$$

$$\langle \bar{\bar{T}}_{\text{GEL}} (\text{in}) \rangle = \langle \bar{\bar{T}}_{\text{GEL2}} (\text{in}) \rangle, (\varepsilon_{\text{Eff}}, \mu_{\text{Eff}}) = (\text{Re}[\varepsilon_s], \mu_0), \text{when extremely absorbing}; \quad (22b)$$

$$\langle \bar{\bar{T}}_{\text{GEL}} (\text{in}) \rangle = \text{unknown}; \quad \text{between moderate and extreme absorption}. \quad (22c)$$

In Eq (22b), $\langle \bar{\bar{T}}_{\text{GEL2}}(\text{in}) \rangle = \frac{1}{2}\text{Re}\left[ D_{\text{in}} E_{\text{in}}^* + B_{\text{in}} H_{\text{in}}^* - \frac{1}{2}\left( \mu_0 H_{\text{in}} \cdot H_{\text{in}}^* + \text{Re}[\varepsilon_s] E_{\text{in}} \cdot E_{\text{in}}^* \right) \bar{\bar{I}} \right]$. The approximate absorption region in which the absorption of a scatterer is much higher than the 'moderate absorption' (previously discussed in section III.A) is defined as the region of 'extreme loss' (i.e. $\varepsilon_I \gg \text{background permittivity}, \varepsilon_b$). We define the internal dynamics of such an object as the 'GEL system of second kind'. Illustration of such a system is shown in Fig. 4s in the supplement S4 (d). In Eq (22b), 'GEL2' denotes the second kind GEL formulation. In such situations, only the appropriate definition of $P_{\text{Eff}}$ and $M_{\text{eff}}$ in Eqs. (7a) and (7b) is required to obtain the total interior force via the GEL theory, (see note [49]). Fig. 4s in Supplement S4 (d) shows that for an 'extremely absorbing' scatterer embedded in air or vacuum, one requires $P_{\text{Eff}} = (\varepsilon_s - \varepsilon_{\text{Eff}})E_{\text{in}} = i\varepsilon_I E_{\text{in}}$ and $M_{\text{Eff}} = M = (\mu_s - \mu_0)H_{\text{in}}$ to satisfy Eq (9b). Therefore we recognize: $\varepsilon_{\text{Eff}} = \text{Re}[\varepsilon_s] = \varepsilon_R$ and $\mu_{\text{Eff}} = \mu_0$ respectively. In consequence, according to our Eqs (8b), (10b) and (11b) for an extremely absorbing object embedded in air or vacuum, the generalized EL ST, force and momentum density take the final form:

$$\langle \bar{\bar{T}}_{GEL2}(in) \rangle = \frac{1}{2} \text{Re} \left[ D_{in} E_{in}^* + B_{in} H_{in}^* - \frac{1}{2} \left( \mu_0 H_{in} \cdot H_{in}^* + \text{Re}[\varepsilon_s] E_{in} \cdot E_{in}^* \right) \bar{\bar{I}} \right], \quad (23)$$

$$\langle f_{GEL2} \rangle = \frac{1}{2} \text{Re} \left[ (i\varepsilon_I E_{in} \cdot \nabla) E_{in}^* + ((\mu_s - \mu_0) H_{in} \cdot \nabla) H_{in}^* - i\omega(i\varepsilon_I E_{in} \times B_{in}^*) + i\omega((\mu_s - \mu_0) H_{in} \times D_{in}^*) \right] \quad (24)$$

$$G_{GEL2}(in) = \left[ G_{Material} + G_{Electromag} \right] = [\varepsilon_{Eff} \mu_{Eff} (E_{in} \times H_{in}) - (P_{Eff} \times M_{Eff})] = (\text{Re}[\varepsilon_s] \mu_s + \varepsilon_s \mu_0 - \varepsilon_s \mu_s)(E_{in} \times H_{in}) \quad (25)$$

According to Eq (25), the interior of such an extremely lossy object constitutes a 'GEL system', where the total non-mechanical momentum density is neither Abraham's nor Minkowski's. On the other hand, the reason why in the range between the moderate and extreme absorption (cf. Eq (22c)), the GEL2 equations or any other ST cannot yield the total force excluding hidden momenta (and then how the total volume force can be calculated) is discussed in details in Supplement S4 (d).

However, for any generic highly absorbing scatterer embedded in air (cf. Eqs (22b) and (22c)), it is clear that the non-mechanical momentum density at those highly absorption regions is not the one of Abraham, (since EL and Chu ST do not yield the total internal force), neither the one of Minkowski, (since the Minkowski ST does not yield the total internal force); [cf. Fig.4s in Supplement S4 (d)].

### VII. A unified theory: How much unification is possible?

Considering all the above cases, we conclude that based on the appropriate choice of the *effective* polarization $P_{Eff}$, magnetization $M_{Eff}$, and the GEL equations, it is indeed possible to establish appropriate equations that correctly yield the force inside a generic arbitrary macroscopic object, and also in the embedding background. Nonetheless, in order to achieve a unified theory for the conservation of linear momentum, the connection of the GEL formulation, aimed to interior force determinations, with the 'outside force' on the object should be considered. This correct 'outside force' is given by the equations of note [56], which also reveals the connection of the GEL force with the 'outside force'. Our analysis in note [56] shows that even with $P_{Eff}$ and $M_{Eff}$, a unique universal force law both in the interior and exterior of a scatterer, (like e.g. a Mie object), is not possible. But the ST for the 'outside force' still remains Minkowski's [36] along with $p_{\text{Mink}}(\text{out})$ for the momentum transferred to the embedded generic object. This unification can be achieved with $P_{\text{Eff}} = M_{\text{Eff}} = 0$ in our GEL ST and momentum density [cf. Eqs (8a) and (11a)] by expressing them in terms of the outside fields of an embedded generic scatterer, i.e. the GEL Eqs (8a) and (11a) take on the final form: $\bar{\bar{T}}_{\text{GEL}}^{\text{out}} = \bar{\bar{T}}_{\text{Mink}}^{\text{out}}$ and $G_{\text{GEL}}(\text{out}) = G_{\text{Mink}}(\text{out})$.

However, for the experiment reported in [29], it is considered in ref. [4] that the 'outside force' supports the EL formulation. In this connection one should notice that the difference between the 'outside dipole force' [36] and the outside GEL force discussed in note [56] mainly arises from the $\langle F_{\text{e-m}}^{\text{out}} \rangle$ term. If the dipolar object is a dielectric embedded in another dielectric, the total dipole force equation [57]

$$F_{\text{Dipole}}^{\text{out}}(\text{Total}) = (d_e \cdot \nabla)E + \frac{\partial d_e}{\partial t} \times B$$ and the EL force density $f_{\text{EL}} = (P \cdot \nabla)E + \frac{\partial P}{\partial t} \times B$ look similar since

$\langle F_{\text{e-m}}^{\text{out}} \rangle = 0$. But the induced electric dipole moment $d_e = \alpha_e E_{\text{out}}$, (where $\alpha_e$ is the complex electric polarizability defined in [36]) in the outside dipole force equation [57] is applicable only at the exterior of the dipolar object, (cf. Fig.1 in [58]). On the other hand, the polarization $P = (\varepsilon_s - \varepsilon_0)E_{\text{in}}$ in the EL equation [28], or the effective polarization $P_{\text{Eff}} = (\varepsilon_s - \varepsilon_b)E_{\text{in}}$ in the first kind GEL equation (17) is applicable at the interior of any generic macroscopic scatterer such as Rayleigh, dipole, Mie, or even a more complex body. The time averaged EL force constitutes the total internal force on a macroscopic object, which is connected with the concept of force distribution and 'delivered momentum' inside a generic macroscopic object. In contrast, $\langle F_{\text{Dipole}}^{\text{out}}(\text{Total}) \rangle$ simply describes the 'outside force', [cf. Eq (5a)], which is connected with the concept of 'transferred momentum' from the background to a dipole.

These aformentioned differences can be verified in a simple way: if for example the 'outside force' on a dipolar dielectric sphere (embedded in another dielectric) is calculated based on the EL ST (rather than on Minkowski ST) by employing only the background fields, it will lead to an incorrect result. $\langle F_{\text{Dipole}}^{\text{out}}(\text{Total}) \rangle$ and the time-averaged 'outside force' obtained from the EL ST do not match because the stress tensor and the photon momentum associated with the total dipole force [57] is nothing else but the one of Minkowski [36]. In consequence, although $F_{\text{Dipole}}^{\text{out}}(\text{Total}) = (d_e \cdot \nabla)E + \frac{\partial d_e}{\partial t} \times B$ [57] looks like the EL force, the 'outside force' used in [29] is not the EL one. Rather it is the force $\langle F_{\text{Dipole}}^{\text{out}}(\text{Total}) \rangle$ of Eq (N56a) in our note [56], which is associated only with the external Minkowski ST Eq (5b) [36]. It is notable that our formulation pertains to the dynamics of macroscopic objects, which is different to that of a single atom [59, 60] or other microscopic entities [48].

**VIII. Key conclusions on the 'existence domain' of the different stress tensors and force laws**

(1) In section III.A, and in Supplements S0 and S1, we clarify that if one determines the internal force on an object with no or moderately loss embedded in air or vacuum, $G_{\text{Abr}}$ appears supporting only the Einstein-Laub equations, [cf. Eq (3)] by excluding hidden momenta [see Figs.1s (a) and (b)]. Thus only such a scatterer can be modelled as a transparent Einstein-Balazs box supporting the internal photon momentum $p_{\text{Abr}}(\text{in})$ [cf. Eq (4)].

(2) If the absorption of a scatterer embedded in air approximately reaches the extreme loss (which is much higher than the moderate loss of a scatterer), its internal force can be calculated excluding hidden momenta by GEL2 ST in Eq (23). Between the range of moderate and extreme absorption, it is not possible to yield the total internal force on the scatterer excluding hidden momentum. However, if the absorption of a body embedded in air crosses the moderate loss, its internal non-mechanical momentum density does not remain as the one of $G_{\text{Abr}}(\text{in})$.

(3) If the background is not air or vacuum, the internal force on an embedded object cannot be calculated from the EL equations (3); then our first kind GEL equations (14) and (17) should be used. The 'outside force' on the embedded object should be obtained by Minkowski's ST [cf. Eqs (5a) and (5b)]. Thus at least two different STs (i.e. external and internal) determine the total force independently of the object by excluding hidden momentum. In Figs. 3(a) to Fig. 4(b) in this paper, and in Supplement S4 (a)-(d), we have shown several numerical and full wave simulation results that support Eq (19), [or more generally Eq (9b)]. Eq (19) also supports the concept: 'existence domain' for STs, force laws, and photon momenta. Our theory can successfully interpret and explain the different experimental observations reported in [7], [13], [26], [29] and [33]. Moreover, our conclusions remain valid even if an object is embedded in a heterogeneous [cf. Fig. 3(b)] and bounded background [cf. Fig. 4(a)].

(4) There are several experiments that some of the previous STs and optical force equations have failed to interpret because they have been used without considering their proper domains of validity [48,50, 61], (see also our discussion in Section VII). For example, if one considers a rectangular magneto-dielectric slab embedded in a magneto-dielectric medium, the force that applies should be that Eq (20a), which supports Eq (19). To our knowledge, no other ST, or force law, leads to such a matching of the interior force with the outside one without adding hidden quantities. This and our several other results in Figs. 3(a)–4(b), and in Supplement S4 (a)-(d), explain why for several experimental situations [61], the Helmholtz/Minkowski force (cf. Table-1), which is connected with the external Minkowski ST and $p_{\text{Mink}}$ (see Table-1), with outside fields leads to the correct results. However, if this same Helmholtz force is applied inside an object, it may lead to a wrong result or create some serious ambiguities, even for some simple cases as reported in [50].

(5) Previously the experiment of [34] and the photon drag experiment of [53] caused a severe controversy [2-4, 20, 35] regarding the force and photon momentum in the dielectric background [34] or in the host [53], respectively. To solve those dilemmas, based on our concept: 'existence domain' we conclude that if one measures the internal force on the large background excluding hidden momenta, the appropriate formulation should support the EL equations with the background fields and with $p_{\text{Abr}}(\text{in})$ ,[cf. Eq (21) and Sections V. A and V. B]. In contrast, if one measures the transfer of momentum to the scatterer, or scatterers, embedded in that background or host, the appropriate formulation should support

Minkowski's equations with the background, or host, fields, i.e. with the external fields of the embedded scatterer; [cf. Eq (5a) and (6)].

Thus summarizing all the above, we state that our Eq (19) supports the 'existence domain' of Minkowski's ST and our first kind GEL excluding hidden quantities. We have verified the validity of Eq (19) [or, more generally, of Eq (9b)] based on several numerical results via full Mie calculations along with full wave simulation results, [se Figs. 3(a) - 4(b) and Supplements S0 and S4 (a)-(d)]. These all, along with Table-1, should finally resolve most previous conflicts reported in [2-4, 13-24, 26-29, 33-42,48,50,51-53,59]. Especially the compact classification provided in Figs. 1(a)-(c) should allow the prediction of the correct 'measurent approach' of the total force, excluding hidden quantitites, as well as the photon momenta, for the different situations without further ambiguities.

## IX. Key conclusions on the 'existence domain' of the different photon momenta

(1)  The resolution of the dilemma of two photon momenta in the same background medium can be well understood by recognizing the completely different roles of $\boldsymbol{p}_{\text{Abr}}(\text{in})$ and $\boldsymbol{p}_{\text{Mink}}(\text{out})$ in Eqs. (21) and (6), respectively. If a measurement is done only to determine the transfer of optical momentum into a body (i.e. into the embedded scatterer), Minkowski photon momentum should emerge at its interfaces, or boundaries. By contrast, if a measurement is done to get the appropriate internal force and the force distribution of the embedding background, the pure electromagnetic photon momentum in that large continuous background should emerge as that of Abraham. Interestingly, this conclusion unifies four different confusing experiment interpretations [7, 34, 52, 53] in a compact framework.

(2)  *The GEL theory answers what the behavior of the photon momentum inside the scatterer should be.* For example, if one measures the total internal force, or force distribution, on a scatterer embedded in a non-vacuum background, the associated photon momentum inside this body should emerge as $\boldsymbol{p}_{\text{Non-Mech.}}(\text{in})$ defined in Eq (16), where $p_{\text{Non-Mech.}}(\text{in}) = p_{\text{Material}}(\text{in}) + p_{\text{Abr}}(\text{in})$. Considering Eqs (15) and (23), one of the main conclusions in this work is: the GEL non-mechanical momentum density is indeed a function of our newly defined $P_{\text{Eff}}$ and $M_{\text{Eff}}$ inside the object, [cf. Eq (11a)], and coincides with Abraham's momentum density only for moderately or non-absorbing scatterers embedded in air.

(3) Inside an embedded object a Lagrangian description, along with a conventional electric and magnetic dipole moment [24, 59, 62], always leads to: $p_{\text{Cano.}}^{\text{med}} + p_{\text{Mink}} = p_{\text{kin}}^{\text{med}} + p_{\text{Abr}}$. However, any ST associated with $p_{\text{Abr}}(\text{in})$ does not lead to a correct time-averaged interior force on the embedded object. Moreover, an internal Minkowski ST (i.e. by employing the fields inside the object) associated with $p_{\text{Mink}}(\text{in})$ leads to a zero time-averaged force on a lossless object, as nothing is embedded inside it. As a result, according to our verifications in Fig. 3(a) - 4(b) and Supplement S4 (a)-(d) on the validity of the first kind GEL ST (Eq (14)) and force law ( Eq (17)), the appropriate equation of total momentum conservation inside an embedded object must be Eq (16). I.e., according to Eqs (15) and (16) we have: $p_{\text{Non-Mech.}}(\text{in}) \neq p_{\text{Abr}}(\text{in})$ and $p_{\text{Non-Mech.}}(\text{in}) \neq p_{\text{Mink}}(\text{in})$.

(4) As stressed in Section V A, *the dilemma on the appropriate photon momentum is not about its general correctness, but on which one emerges in a specific measurement*. The answer to this question is contained in the three previous points above in this Section. The *'existence domains'* of the electromagnetic momentum densities (or photon momenta) in above three points, as well as the proposal of kinetic and canonical momenta reported in [24], can be unified via the equations of total momentum conservation: (4), (6), (13) and (21); even though Eq (4) is a special case of Eq (13) [or of Eq (16)]. Hence, although the transfer of momentum from the background to an embedded macroscopic object can be considered as $p_{\text{Mink}}(\text{out})$, [specially at the boundary of the object], the internal non-mechanical momentum inside the object can achieve different magnitude other than that of $p_{\text{Abr}}(\text{in})$ and $p_{\text{Mink}}(\text{in})$.

Thus summarizing all the above, a main conclusion is: *the observation of a specific photon momentum dependens on the 'measurement approach'*, which in fact arises from the 'existence domain' of the corresponding ST and photon momentum. This is supported by our main equations (9b), (19), as well as by Eqs. (4), (6), (13), (21). It seems that such a simple interpretation as this one based on the concept of 'existence domain' has remained unnoticed in previous works [2-4,20 ,35,45,48], thus hindering an explanation of the experiments of [2-4,7,20,34,52,53].

## X. Final remarks

In 1918 Albert Einstein himself wrote against his stress tensor proposed with Jacob Laub [63]:

*"It has long been known that the values I had derived with Laub at the time are wrong; Abraham, in particular, was the one who presented this in a thorough paper. The correct strain tensor has incidentally already been pointed out by Minkowski"*

Though apparently not noticed by Einstein at the time, the EL formulation of the force and stress tensor, along with Abraham's photon momentum is suitable inside objects with moderate absorption, embedded in air or vacuum. This is one of our conclusions in this work. In addition, we have shown that if an object is embedded in a material background, the 'outside force' is determined from the background fields on using Minkowski's stress tensor which supports the transfer of momentum $p_{Mink}$ from the background to the embedded object. This can also be done by setting $P_{Eff} = M_{Eff} = 0$ in our Generalized Einstein-Laub ST and momentum density. By the same token, the 'interior force' is determined from the fields inside the body on using our Generalized Einstein-Laub equations, which in turn are in agreement with the 'outside force' given by Minkowski's ST. Nonetheless, in contrast with the 'outside force', the 'interior force' requires an appropriate choice of $P_{Eff}$ and $M_{Eff}$ in the Generalized Einstein-Laub equations. Then according to the linear momentum continuity equation: $\boldsymbol{f} = \nabla \cdot \overline{\overline{\boldsymbol{T}}} - \frac{\partial}{\partial t}\boldsymbol{G}$, if one identifies the appropriate stress tensor, it leads to: (i) The correct time-averaged 'interior force' without any problematic manual addition of hidden quantities. (ii) The correct non-mechanical momentum. This reveals that Abraham's momentum density is only a special case for an arbitrary embedded object, since a more general interior momentum density $G$ should be expressed in terms of our newly defined quantities $P_{Eff}$ and $M_{Eff}$.

We thus conclude that excluding hidden momenta, we require at least two different STs and electromagnetic momentum densities to establishing a linear momentum continuity equation. These are: *Minkowski's ST and momentum density outside the body*, and *our Generalized Einstein-Laub ST and momentum density inside*. This is not only simple for calculations, but more effective than approximate force formulae, such as e.g. the dipolar force [13] approach, to verify and predict the outcome of several complex experimental situations.

Summarizing, in this paper we have demonstrated that a unique force law or ST, along with a unique photon momentum, cannot provide a linear momentum conservation law, since the physical effects derived from the different STs, electromagnetic momentum densities, and force laws are quite different from each other due to their different 'existence domains'. However, although a unique general equation for both the interior and exterior forces law is unfeasible, an expression of the GEL ST and the momentum density is attainable in terms of the vectors $P_{Eff}$ and $M_{Eff}$ defined in this paper.

Hence, if we exclude the problematic hidden momenta, the conservation law: $f = \nabla \cdot \bar{\bar{T}} - \frac{\partial}{\partial t} G$ can be expressed by a single equation based on our postulate of note [49] as:

$$\langle F_{Total} \rangle (\text{out}) \approx \langle F_{Total} \rangle (\text{in}); \text{ Where } \bar{\bar{T}}_{GEL} = \bar{\bar{T}}_{Mink} + \frac{1}{2}(M_{Eff} \cdot H_{in} + P_{Eff} \cdot E_{in})\bar{\bar{I}} \text{ and}$$

$G_{GEL}(\text{in}) = [[(D_{in} - P_{Eff}) \times (B_{in} - M_{Eff})] - (P_{Eff} \times M_{Eff})]$, (for e.g. a sphere or cylinder $\mathbf{P}_{Eff} = \mathbf{M}_{Eff} = 0$ for $r > a$, but $\mathbf{P}_{Eff} = \chi_e \mathbf{E}_{in}$ and $\mathbf{M}_{Eff} = \chi_m \mathbf{H}_{in}$ for $r < a$). $\chi_e$ and $\chi_m$ denote the electric and magnetic susceptibilities, which are respectively functions of the scattering object permittivity and permeability for the different configurations at hand. For example:

(i) When $P_{Eff} = (\varepsilon_S - \varepsilon_b)E_{in}$ and $M_{Eff} = (\mu_S - \mu_b)H_{in}$, $\bar{\bar{T}}_{GEL}^{in} = \bar{\bar{T}}_{GEL1}^{in}$ and $G_{GEL} = G_{GEL1}^{in}$, if the background is air or vacuum, the GEL equations turn into those of Einstein-Laub.

(ii) When $P_{Eff} = i\varepsilon_I E_{in}$ and $M_{Eff} = M = (\mu_S - \mu_0)H_{in}$, then $\bar{\bar{T}}_{GEL} = \bar{\bar{T}}_{GEL2}^{in}$ and $G_{GEL} = G_{GEL2}^{in}$.

(iii) For embedded chiral scatterers, the form of $\chi_e$ and $\chi_m$ will be different, but the final form of the ST derived from the GEL ST, Eq (8b), remains the same as in Eq (14).

However although a single internal time-averaged force expression [i.e. GEL force, cf. Eq (10a)] remain valid in terms of $P_{Eff}$ and $M_{Eff}$ most of the times, it is not possible to achieve unique mathematical expression that includes both the inside and outside force [56].

Our study also resolves the dilemma on two possible photon momenta in an embedding background by identifying the two fully different concepts based on the measurement approach: 'mementum transfer' and 'delievered momentum'. In particular the possible riddles arising from the experiments of [7], [26], [29], [34], [52] and [53] are sorted out by our unified theory. In this way, we resolve the long-lasting debates between the formulations of Abraham, Einstein-Laub and Minkowski, and unify them with the GEL ST and non-mechanical momentum density formulations. Our theory, that excludes hidden momenta, and embraces the Minkowski and GEL stress tensors as the general calculation entities, should bear consequences not only in physics, but also for engineering and biology areas requiring the determininination and control of optical forces [5, 6, 30-32].

**Acknowledgments**

CWQ is supported by NUS via Grant R-263-000-678-133. MN-V is supported by the Spanish Ministerio de Economia y Competitividad through FIS2012-36113-C03-03 and FIS2014-55563-REDC research grants. We thank Remi Carminati and José A. Sanchez-Gil for valuable comments and a critical reading of the manuscript.

**Figures and Captions**

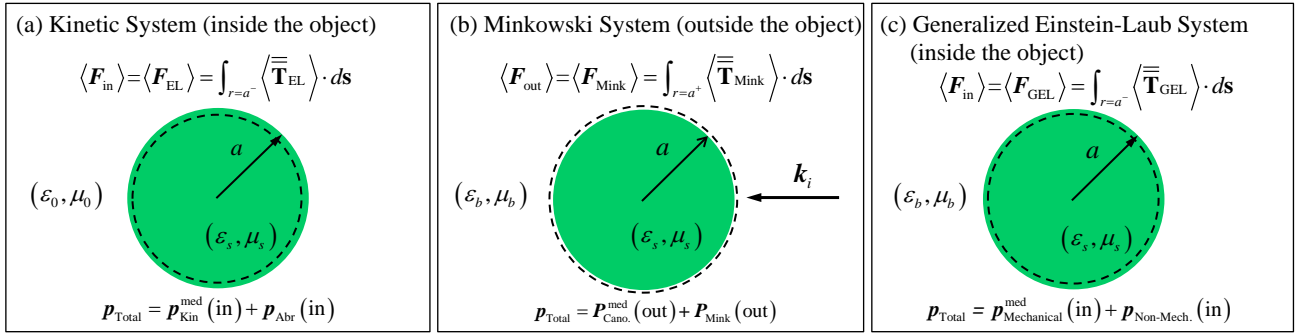

Fig.1: (a) *'Kinetic system'*: A homogeneous lossless, or moderately absorbing, object is embedded in air or vacuum. Inside the object this system supports Eq (1) as the governing equation of total momentum conservation. This leads to the total force, $\langle F_{in} \rangle$ given by the EL equations employing only the interior fields. No other stress tensor except the EL ST excluding hidden momenta leads to total interior force. (b) *'Minkowski system'*: An object is immersed in an arbitrary material background. Outside the body this system supports the transfer of $p_{Mink}(\text{out})$ from one medium to another, which is correctly described by the total momentum conservation Eq (2). The transfer of $p_{Abr}(\text{out})$ is not detectable by any calculation based on the photon momentum such as a ray tracing method or Doppler shift effect. In order to get the total 'outside force' (see the arrows), $\langle F_{out} \rangle$ is found via Minkowski's ST. (c) A *'generalized Einstein Laub (GEL) system'*: A homogeneous object is immersed in an arbitrary medium. Inside this body, such systems support our equation for the total momentum conservation: $p_{Total} = p^{med}_{Mechanical}(\text{in}) + p_{Non\text{-}Mech.}(\text{in})$, which is a generalization of Eq (1), (see Eq. (16) of text and section III.C.b). This leads to the total force $\langle F_{in} \rangle$ inside the object given by our Generalized Einstein-Laub equations (GEL). No other stress tensor except the GEL one leads to correct interior force excluding hidden momenta.

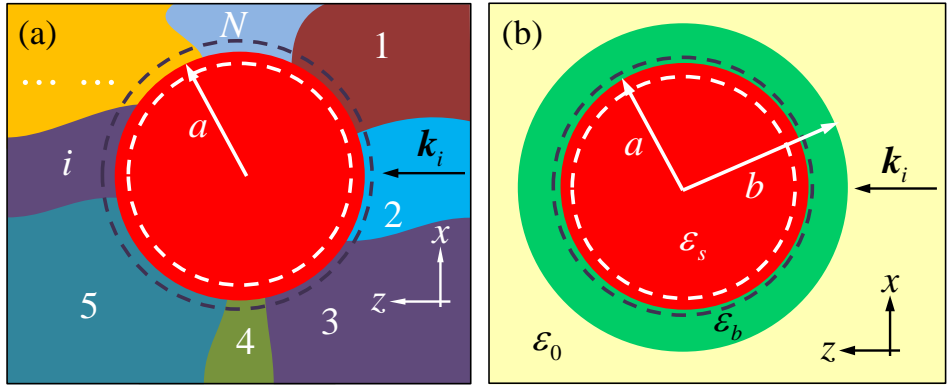

Fig. 2: Illustration of momentum transfer and delivery from an incident plane wave with propagation vector $k_i$ to an embedded three-dimensional (3D) generic object, (i.e. Rayleigh, dipole, Mie, or larger than Mie, object). (a) A sphere or cylinder is immersed in an unbounded and heterogeneous background medium. (b) A core-shell sphere or cylinder (i.e., the core is embedded in a bounded background). In both cases the total force obtained by using the time-averaged ST is $\langle F_{out} \rangle$ evaluated from fields outside the particle ('Minkowski system'), at $r = a^+ = 1.001a$, (black circles); whereas this force is $\langle F_{in} \rangle$ when the ST is determined from fields inside the object ('GEL system') at $r = a^- = 0.999a$ (white circles). Almost all the previous conflicts of linear momentum conservation arise from the inappropriate choice of STs, (or force laws), and photon momenta. It is for this reason that the 'existence domain' of the different STs (and force laws without hidden quantities) has not been previously identified, (see Table 1). By excluding hidden momenta, we show that $\langle F_{out} \rangle$ should be calculated from Minkowski ST, whereas $\langle F_{in} \rangle$ should be obtained by our GEL ST. If and only if the background is air or vacuum, $\langle F_{in} \rangle$ should be given by the EL ST, which is a special case of our GEL ST. Other stress tensors do not match.

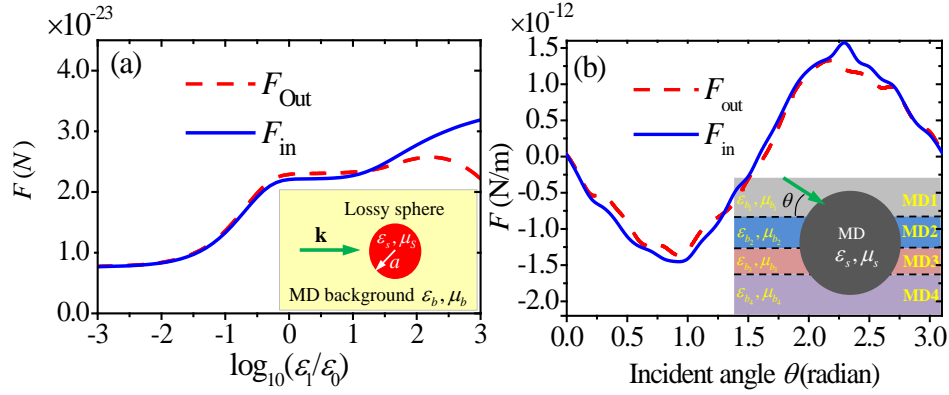

Fig.3: Time-averaged forces ($F_{out}$ at $r = a^+ = 1.001a$ from Minkowski ST, Eq (5), and $F_{in}$ at $r = a^- = 0.999a$ from the first kind GEL (GEL1) ST, Eq. (14). These forces are in accordance with Eq (19). (a) Force on an absorbing magneto-dielectric sphere with $a$=1000nm (i.e., a Mie object), $\varepsilon_s = 15\varepsilon_0 + i\varepsilon_I$, $\mu_s = 3\mu_0$ illuminated by a linearly polarized plane wave $E_x = e^{i(kz-\omega t)}$ at $\lambda = 1064$ nm. The unbounded homogeneous magneto-dielectric background parameters are: $\varepsilon_b = 4\varepsilon_0$ and $\mu_b = 2\mu_0$. Notice that the sphere loss is in log scale. (b) Force on a magneto-dielectric infinite cylinder of ($\varepsilon_s, \mu_s$) = ($5\varepsilon_0$, $2\mu_0$) and radius *2000 nm* embedded in heterogeneous unbounded background of four different magneto-dielectric layers: ($\varepsilon_b, \mu_b$) = ($3\varepsilon_0$, $2\mu_0$); ($4\varepsilon_0$, $3\mu_0$); ($5\varepsilon_0$, $4\mu_0$); ($6\varepsilon_0$, $5\mu_0$) [cf. also Supplement S4(c)]. The illuminating plane wave with $\lambda = 1064$ nm incides at varying angles.

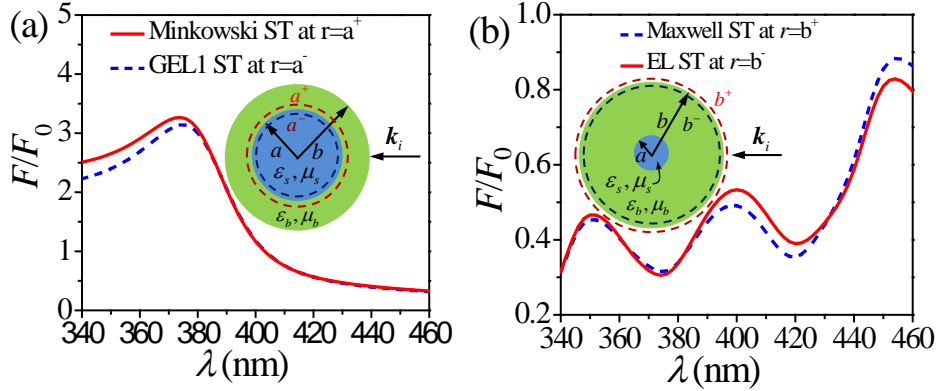

Fig. 4: A 3D core-shell dielecrtic sphere embedded in air is illuminated by a linearly polarized plane wave. (a) Core radius, $a$=50 nm, $\varepsilon_s = (15+0.1i)\varepsilon_0$, $\mu_s = \mu_0$. Bounded local immediate background (i.e. the shell) parameters: radius, $b$=60 nm and $\varepsilon_b = 4\varepsilon_0$; $\mu_s = \mu_0$. $\langle F_{out}^{Core} \rangle$, in Newtons, at different illumination wavelengths $\lambda$ obtained from Minkowski ST at $r=a^+$ using the fields in the shell. Force $\langle F_{in}^{Core} \rangle$ based on the first kind GEL (GEL1) ST [Eq. (14)] at $r=a^-$ using core fields. $F_0$ is the scaling factor: $F_0 = 5 \times 10^{-26}$ N. This adequately described by Eq (19). (b) A core-shell sphere with $b>>a$, [$b=8a$] is embedded in air. Core radius $a$=50 nm, $\varepsilon_s = (15+0.1i)\varepsilon_0$, $\mu_s = \mu_0$; $\varepsilon_b = 4\varepsilon_0$ and $\mu_b = \mu_0$. Spectra of $\langle F_{out}^{Shell} \rangle$ calculated by the Maxwell ST at $r=b^+$ using fields in air, and of $\langle F_{in}^{Shell} \rangle$ given by the EL ST, at $r=b^-$ with fields in the entire object. $F_0 = 2.23 \times 10^{-24}$ N. $\langle F_{in}^{Shell} \rangle$ can be considered as the 'force felt' by the background (i.e. the large shell).

## References and notes

[1]    J. Ng, J. Chen, Z. Lin, and C.T. Chan, Optical puling force. Nat. Photonics **5**, 531 (2011).

[2]    C. Baxter, and R. Loudon, Radiation pressure and the photon momentum in dielectrics. J. Mod. Opt. **57**, 830 (2010).

[3]    D.J. Griffiths, Electromagnetic Momentum. Am. J. Phys. **80**, 7 (2012).

[4]    B.A. Kemp, Resolution of the Abraham-Minkowski debate: Implications for the

electromagnetic wave theory of light in matter. J. Appl. Phys. **109**, 111101 (2011).

[5]     A. Ashkin and J.M. Dziedzic, Optical trapping and manipulation of viruses and bacteria. Science **235,** 1517 (1987).

[6]     M.C. Zhong, X.B. Wei, J.H. Zhou, Z.Q. Wang, and Y.M. Li, Trapping red blood cells in living animals using optical tweezers. Nat. Commun. **4**, 1768 (2013).

[7]     V. Kajorndejnukul, W. Ding, S. Sukhov, C.-W. Qiu, and A. Dogariu, Linear momentum increase and negative optical forces at dielectric interface. Nat. Photonics **7**, 787 (2013).

[8]     J.R. Moffitt, Y.R. Chemla, S.B. Smith, and C. Bustamante, Recent Advances in Optical Tweezers. Annu. Rev. Biochem. **77**, 205 (2008).

[9]     V. Shvedov, A. Davoyan, C.R. Hnatovsky, N. Engheta , and W.A Krolikowski, long-range polarization-controlled optical tractor beam. Nat. Photonic*s* **8**, 846 (2014).

[10]    A. Novitsky, C.-W. Qiu, and H. Wang, Single gradientless light beam drags particles as tractor beams. Phys. Rev. Lett. **107,** 203601 (2011).

[11]    M. Li, W.H.P. Pernice, and H. X. Tang, Tunable bipolar optical interactions between guided lightwaves.  Nat. Photonics **3**, 464 (2009).

[12]    G.S. Wiederhecker, L. Chen,  A. Gondarenko, and M. Lipson, Nature **462**, 633 (2009) and A. W. Rodriguez, P.Ch. Hui, D.N. Woolf, S.G. Johnson, M. Loncar, and F. Capasso, Classical and fluctuation-induced electromagnetic interactions in microscale systems: designer bonding, antibonding and Casimir forces, arXiv:1409.7348 **[**physics.optics] (2014).

[13]    M.L. Juan, R.Gordon, Y. Pang, F. Eftekhari and R. Quidant, Self-induced back-action optical trapping of dielectric nanoparticles. Nat. Phys. **5**, 915 (2009).

[14]    M. Mansuripur, Optical manipulation: Momentum exchange effect. Nat. Photonics **7**, 765 (2013).

[15]    R. Loudon, L. Allen, and D.F. Nelson, Propagation of electromagnetic energy and momentum through an absorbing dielectric. Phys. Rev. E **55**, 1071 (1997).

[16]    R. Peierls, The Momentum of Light in a Refracting Medium. Proc. R. Soc. London Ser. A **347**,

475 (1976).

[17]   B.A. Kemp, T.M. Grzegorczyk, and J.A. Kong, Optical Momentum Transfer to Absorbing Mie Particles. Phys. Rev. Lett. **97**, 133902 (2006).

[18]   B.A. Kemp, J.A. Kong, and T.M. Grzegorczyk, Reversal of wave momentum in isotropic left-handed media. Phys. Rev. A **75**, 053810 (2007).

[19]   K.J. Webb and Shivanand. Negative electromagnetic plane-wave force in gain media. Phys. Rev. E **84**, 057602 (2011).

[20]   K.J. Webb, Dependence of the Radiation Pressure on the Background Refractive Index. Phys. Rev. Lett. **111**, 043602 (2013).

[21]   M. Mansuripur, Trouble with the Lorentz Law of Force: Incompatibility with Special Relativity and Momentum Conservation. Phys. Rev. Lett. **108**, 193901 (2012).

[22]   A. Einstein and J. Laub, Annalen der Physik **331**, 541 (1908); The English translation of this paper appears in Einstein's Collected Papers, Vol. 2, Princeton University Press (1989).

[23]   K.T. Mcdonald, Biot-Savart vs. Einstein-Laub Force Law. Joseph Henry Laboratories, Princeton University, Princeton, NJ 08544, (2013), http://www.physics.princeton.edu/~mcdonald/examples/laub.pdf.

[24]   S.M. Barnett, Resolution of the Abraham-Minkowski Dilemma. Phys. Rev. Lett. **104**, 070401 (2010).

[25]   We reamark the similarity between our concept: 'existence domain' and the idea of local property in quantum mechanics. According to Y. Aharonov [cf. Y. Aharonov, *Aharonov-Bohm effect and non local phenomena*, Proc. Int. Symp. Foundations of Quantum Mechanics and their Technical Implications, Tokyo, 10 (1984)]: "The essential difference between local and non-local properties of the system is that in the former case all possible information can be obtained by *independent measurements made in the two regions*, while in the latter case this is not true."

[26]   J. Guck, R. Ananthakrishnan, T. J. Moon, C. C. Cunningham, and J. Kas, Optical Deformability of Soft Biological Dielectrics. Phys. Rev. Lett. **84**, 5451 (2000).

[27]     A. Zakharian, M. Mansuripur, and J.V. Moloney, Radiation pressure and the distribution of electromagnetic force in dielectric media. Opt. Express **13**, 2321 (2005).

[28]     M. Mansuripur, A.R. Zakharian, and E.M. Wright, Electromagnetic force distribution inside matter. Phys. Rev. A **88**, 023826 (2013).

[29]     A. Rohrbach, Stiffness of optical traps: Quantitative agreement between experiment and electromagnetic theory. Phys. Rev. Lett. **95**, 168102 (2005).

[30]     K. Svoboda, Biological Applications of Optical Forces. Annu. Rev. of Biophys. Biomem. **23**, 247 (1994).

[31]     M.M. Rahman, A.A. Sayem, M.R.C. Mahdy, M.E. Haque, R. Islam, S.T. R. Chowdhury, M. Nieto-Vesperinas, M.A. Matin, Material Independent Long Distance Pulling, Trapping, and Rotation of Fully Immersed Multiple Objects with a Single Optical Set-up. ArXiv: 1504.00638 [physics.optics] (2015).

[32]     M. Hoeb, J.O. Radler, S. Klein, M. Stutzmann, and M.S. Brandt, Light-induced dielectrophoretic manipulation of DNA. Biophys J. **93**, 1032 (2007).

[33]     C.-W. Qiu, W. Ding, M.R.C. Mahdy, D. Gao, T. Zhang, F.C. Cheong, A. Dogariu, Z. Wang and C.T. Lim, Photon momentum transfer in inhomogeneous dielectric mixtures and induced tractor beams. Light: Science & Applications **4**, e278 (2015).

[34]     R.V. Jones and B. Leslie, The measurement of optical radiation pressure in dispersive media. Proc. Roy. Soc. A **360**, 347 (1978).

[35]     B.A. Kemp, and T.M. Grzegorczyk, The observable pressure of light in dielectric fluids. Opt. Lett. **36**, 493 (2011).

[36]     M. Nieto-Vesperinas, J.J. Sáenz, R. Gómez-Medina, and L. Chantada, Optical forces on small magnetodielectric particles. Opt. Express **18**, 11428 (2010).

[37]     L.P. Pitaevskii, Comment on "Casimir force acting on magnetodielectric bodies embedded in media". Phys. Rev. A **73**, 047801 (2006).

[38]     I. Brevik, and S.A. Ellingsen, Comment on "Casimir force acting on magnetodielectric

[39]     B.A. Kemp, T. M. Graegorczyk, and J. A. Kong, Ab initio study of the radiation pressure on dielectric and magnetic media. Opt. Express **13**, 9280 (2005).

[40]     M. Mansuripur, and A.R. Zakharian, Whence the Minkowski momentum? Optics Communications **283**, 3557 (2010).

[41]     C.Raabe, and D.-G.Welsch, Casimir force acting on magnetodielectric bodies embedded in media. Phys. Rev. A, **71**, 013814 (2005).

[42]     V.V. Datsyuk and O. R. Pavlyniuk, Maxwell stress on a small dielectric sphere in a dielectric. Phys. Rev. A **91**, 023826 (2015).

[43]     U. Leonhardt, Abraham and Minkowski momenta in the optically induced motion of fluids. Phys. Rev. A **90**, 033801 (2014).

[44]     In the process of optical momentum transfer from the background to an embedded object, the difference between $p_{\text{Mink}}$ (out) and $p_{\text{Abr}}$ (out) lies in the time domain and hence there is an extra term $f_A$ in Abraham's force law, which does not occur in the Helmholtz/Minkowski force (cf. [33], Section: 'validity of other methods', and also Table-1 of this paper). Although the time-averaged Abraham ST, linked to $p_{\text{Abr}}$ (out), gives the same 'outside force' as Minkowski's ST, $p_{\text{Abr}}$ (out) corresponds to a specific situation that only arises in the time domain. As a matter of fact, the photon momentum based approaches, such as ray tracing calculations [7, 33] and Doppler shift effects [4, 35], support $p_{\text{Mink}}(\text{out})$ rather than $p_{\text{Abr}}$ (out) [4].

[45]     From de Broglie's relation, the expression $p_{\text{Mink.}}(=\hbar k = nh/\lambda = nhf/c = np_0)$ can be considered as a measurable momentum that shows three important charachteristics of the canonical momentum, $p_{\text{cano}}$, (cf. E. Leader et al., Phys. Rep. **541**,163 (2014); $\lambda$ and $f$ are the wavelength and frequency of light inside the object of refractive index $n$). $p_{\text{Mink}}$ is connected with the 'transfer of wave momentum'. In contrast, $p_{\text{Abr.}}(\text{in})[=mv=(E/c^2)(c/n)=hf/nc=p_0/n]$ has been defined as a field

momentum [4] or kinetic photon momentum [24], which is a purely non-mechanical part of the wave momentum ($p_{Mink}$) [4]. We have shown that $p_{Abr}$ is connected with the force distributed inside the object, (i.e. it corresponds to a 'kinetic system').

[46] For the purpose of analysing the optical moemtum transfer from the background, one may write [33]: $p_{Mink}(out) = p_{Abr}(out) + p_{med}(out)$. Since $p_{Mink}(out)$ is considered as the canonical momentum and $p_{Abr}(out)$ as the kinetic momentum: $mV$ [45] of photons; the $p_{med}(out)$ term that arises on the boundaries [33, 40] due to the reduced impedance mismatch [40] or Doppler shift effect [48], can be considered as the classical analogue of the '$qA$' term in the relationship: $p_{cano} = mV + qA$ [48]. $m$ is the mass of the particle, $V$ its velocity, $q$ the electric charge, and $A$ the vector potential.

[47] The $p_{med}(out)$ term arises near the boundary from the extra Abraham term, $f_A$, (cf. Table-1 and notice that $p_{med}(out) = \int (n_b^2 - 1) \left( \frac{E_{out} \times H_{out}}{c^2} \right) dv$ and $f_A(out) = (n_b^2 - 1) \frac{\partial}{\partial t} \left( \frac{E_{out} \times H_{out}}{c^2} \right)$). The $f_A$ term, which accounts for the mechanical momentum transfer from the background to the embedded body, vanishes when the time-average is performed in the Abraham force because it is carried by the field. $f_A$ actually hides inside the Minkowski momentum density, $G_{Mink}(out)$ according to the linear momentum conservation equation: $\frac{\partial p_{Total}}{\partial t} = f_{Abr} + \frac{\partial}{\partial t} G_{Abr} = [f_{Mink} + f_A] + \frac{\partial}{\partial t} G_{Abr} = f_{Mink} + \frac{\partial}{\partial t} G_{Mink}$, (cf. Table-1), which supports the conservation of total momentum at any instant [24]: $p_{Total} = p_{kin}^{med} + p_{Abr} = p_{Cano.}^{med} + p_{Mink}$. This interpretation, crucial to understand the important role of Eq (2) in radiation pressure experiments, has been overlooked in previous works. This is the reason why direct photon momentum based approaches (such as ray tracing procedures, Doppler Shift effect and also Eq (4) in [36]) are always in agreement with the transfer of the momentum $p_{Mink}(out)$.

[48] P.W. Milonni, and R.W. Boyd, Momentum of Light in a Dielectric Medium. Adv. Opt. Photon. **2**, 519 (2010).

[49] Our postulate is: "*The applicable final form of the STs and electromagnetic momentum densities of the different 'systems' should be expressed independently of $P_{Eff}$ and $M_{Eff}$, although their respective formulations should start with $P_{Eff}$ and $M_{eff}$.*". The importance of this postulate can be understood with this example: Let us consider a dielectric object embedded in a dielectric background. Then the calculations using Eq (8a) should start with $P_{Eff} = (\varepsilon_S - \varepsilon_{Eff})E_{in}$ and $M_{Eff} = (\mu_S - \mu_{Eff})H_{in}$ but in the final applicable form one has to use $\varepsilon_{Eff} = \varepsilon_b$ and $\mu_{Eff} = \mu_b = \mu_0$ to get : $\bar{\bar{T}}_{GEL} = D_{in}E_{in} + B_{in}H_{in} - \frac{1}{2}(\mu_{0(j)}H_{in} \cdot H_{in} + \varepsilon_{b(j)}E_{in} \cdot E_{in})\bar{\bar{I}}$. Notice that if one started with $M_{Eff} = 0$ in Eq (8a), (based on the assumption, for example, that both the background and the scatterer are dielectrics where the conventional magnetization is zero), this would lead to an incorrect form of the GEL ST.

[50] W. Frias and A.I. Smolyakov, Electromagnetic forces and internal stresses in dielectric media. Phys. Rev. E **85**, 046606 (2012).

[51] B.A. Kemp, Comment on Revisiting the Balazs thought experiment in the presence of loss: electromagnetic-pulse-induced displacement of a positive-index slab having arbitrary complex permittivity and permeability. Appl. Phys. A **110**, 517 (2013).

[52] A. Ashkin and J. M. Dziedzic, Radiation pressure on a free liquid surface. Phys. Rev. Lett. **30**, 139 (1973).

[53] A. F. Gibson, M. F. Kimmitt, and A. C. Walker, Photon drag in Germanium. Appl. Phys. Lett. **17**, 75 (1970).

[54] For a dielectric body partially immersed in a bi-background, it has been suggested in [33], in order to explain the observations in [7], that $p_{Mink}$ would only account for 'the process of optical momentum transfer' (or emission) into (from) the embedded scatterer rejecting Peierls' and Abraham's approach, although the photon momentum in the large continuous background medium should be considered as that of Abraham. This conclusion in [33] is in agreement with our formulation. Also, our theory is legitimated by the Ashkin and Dziedzic experiment in [52] where the water surface experiences a backward force like the pulling force in [33]. Alhough the momentum transfer to the water surface [40,

46] is considered as $p_{\text{Mink}}$, [52], the force distribution in water supports the EL force favoring $p_{\text{Abr}}$ (in) according to the full wave simulation results in [28]. The idea of generalized or canonical momentum of photon as $p_{\text{Mink.}}$ in Eq (2) of this article (reported in [24]) should be considered as the governing mechanism to describe 'the process of momentum transfer' not only for an embedded object [24] but also for any interface (i.e. to the water surface [52]), which is fully different than 'the process of delivered momentum' (i.e. in the water medium where Abraham's momentum density emerges from EL force [28]).

[55] Our proposal is validated in another major work on radiation pressure, known as the photon drag experiment [53]. Here the free carriers can be considered as the embedded objects, whereas the host dielectric plays the role of the large background. By a fully different approach to ours, the final conclusion regarding photon drag experiment is drawn in S.M. Barnett et al., Phil. Trans. R. Soc. A **368**, 927 (2010) stating: "*The momentum transfer to a body (in our case the charge carriers) within a medium is given by the Minkowski momentum. The momentum of a photon travelling through a host dielectric, however, is given by the Abraham momentum.*" This statement is also partially in agreement with our proposal. We attribute the reason of such observation on 'measuremnt approach' that arises due to the 'existence domain' of the corresponding ST and photon momentum.

[56] An amazing fact is that, the 'outside force' on a generic magneto-dielectric dipolar object embedded in a magneto-dielectric background (cf. M.Nieto-Vesperinas et al. [36]) can also be partially expressed in terms of an alternative form of the Generalized Einstein-Laub force by expressing it only with the exterior fields rather than with those inside the scatterer, [cf. Eq (10a)]:

$$\left\langle \boldsymbol{F}_{\text{Dipole}}^{\text{out}}(\text{Total}) \right\rangle = \left\langle \boldsymbol{F}_{\text{GEL}}^{\text{out}}(\text{dipole}) \right\rangle + \left\langle \boldsymbol{F}_{\text{e-m}}^{\text{out}} \right\rangle \qquad (\text{N56a})$$

$$\left\langle \boldsymbol{F}_{\text{GEL}}^{\text{out}}(\text{dipole}) \right\rangle = \frac{1}{2} \text{Re}\left[ (\boldsymbol{P}_{\text{Eff}} \cdot \nabla) \boldsymbol{E}_{out}^{*} + (\boldsymbol{M}_{\text{Eff}} \cdot \nabla) \boldsymbol{H}_{out}^{*} - i\omega(\boldsymbol{P}_{\text{Eff}} \times \boldsymbol{B}_{out}^{*}) + i\omega(\boldsymbol{M}_{\text{Eff}} \times \boldsymbol{D}_{out}^{*}) \right], \qquad (\text{N56b})$$

$$\left\langle \boldsymbol{F}_{\text{e-m}}^{\text{out}} \right\rangle = -\frac{k_b^4}{3} \sqrt{\frac{1}{\varepsilon_b \mu_b}} \text{Re}[(\boldsymbol{P}_{\text{Eff}} \times \boldsymbol{M}_{\text{Eff}}^{*})] \qquad (\text{N56c})$$

Eqs.(N56a)-(N56c) are applicable if and only if: (a) the dipole approximation remains valid, (cf. Fig.1 in [58]), and (b) $\boldsymbol{P}_{\text{Eff}} = d_e = \alpha_e \boldsymbol{E}_{\text{out}}$ and $\boldsymbol{M}_{\text{Eff}} = \mu_b d_m = \mu_b \alpha_m \boldsymbol{H}_{\text{out}}$ ; $\alpha_e$ and $\alpha_m$ being the complex polarizabilities defined in [36]. $d_e$ and $d_m$ represent the induced electric and magnetic dipole moment respectively. Also, for a dipolar body (N56a)-(N56c) show that the origin of the difference between the $\langle \boldsymbol{F}_{\text{GEL}}^{\text{out}}(\text{dipole}) \rangle$ and the total 'outside force' is mainly due to the $\langle \boldsymbol{F}_{\text{e-m}}^{\text{out}} \rangle$ term. However, for a Mie object, the 'outside force' takes a more complex form [cf. Eq (39), in the supplement of [1]) than the dipole force; and by no means that 'outside force' in [1] is connected with any alternative form of the outside GEL force. However, always (i.e. in either Rayleigh, or Mie, or any other complex object) the appropriate time-averaged external ST and photon momentum remain Minkowski's ST and $\boldsymbol{p}_{\text{Mink}}(\text{out})$.

[57]    P.C. Chaumet and M. Nieto-Vesperinas, Time-averaged total force on a dipolar sphere in an electromagnetic field. Opt. Lett. **25**, 1065 (2000).

[58]    D. Gao, A. Novitsky, T. Zhang, F. C. Cheong, L. Gao, C. T. Lim, B. Luk'yanchuk, and C.-W. Qiu, Unveiling the correlation between non-diffracting tractor beam and its singularity in Poynting vector. Laser Photon.Rev **9**, 75 (2014).

[59]    S.M. Barnett, and R. Loudon, The enigma of optical momentum in a medium. Phil. Trans. R. Soc. A **368**, 927 (2010).

[60]    E. A. Hinds and Stephen M. Barnett, Momentum exchange between light and a single atom: Abraham or Minkowski? Phys. Rev. Lett. **102**, 050403 (2009).

[61]    I. Brevik, Experiments in phenomenological electrodynamics and the electromagnetic energy-momentum tensor. Phys. Rep. **52**, 133 (1979).

[62]    V. Hnizdo, Comment on Electromagnetic force on a moving dipole. Eur. J. Phys. **33** L3 (2012).

[63]    The Collected Papers of A.Einstein, Vol 8, "The Berlin Years: Correspondence, 1914-1918," Item 565,To W. Dällenbach, after 15 June 1918, Princeton University Press, Princeton, New Jersey, 1998.

# Supplementary Information for "A unified theory correcting Einstein-Laub's electrodynamics solves dilemmas in the electromagnetic stress tensors and photon momenta"

M.R.C. Mahdy[1], Dongliang Gao[1], Weiqiang Ding[1], M. Q. Mehmood[1], Manuel Nieto-Vesperinas[2], Cheng-Wei Qiu[1*]

[1]Department of Electrical and Computer Engineering, National University of Singapore, Singapore
[2]Instituto de Ciencia de Materiales de Madrid, Consejo Superior de Investigaciones Cientificas, Campus de Cantoblanco, 28049, Madrid, Spain

21-page supplement article
Supplemental Figures: Figs. 1s–4s.
**Supplemental Information checklist:**

## S-0: Validity of Einstein-Laub's stress tensor for the interior of an object embedded in air:

To accomplishing our correction of Einstein-Laub (EL) theory in order to handle the complex interior dynamics of an embedded scatter, such as a heterogeneous background configuration, [see Fig.2 (a) in the main article], or a bounded background [see Fig.2 (b) in the main paper], we have first to establish the ST and force law associated to 'kinetic systems', i.e. inside bodies embedded in air or vacuum. Three forces and corresponding ST theories exist for a 'kinetic system' supporting the Abraham momentum density: those of Abraham, Chu, and EL [1]. We recall that Minkowski ST with fields inside a body embedded in air [cf. Fig. 1s (a)] only yields the conduction force [2] rather than the total one. Also, it is well known that the time averaged Abraham ST gives the same result as the Minkowski ST inside a non-absorbing homogeneous and isotropic object [2]. However, for a lossless particle, both Abraham ST and Minkowski ST lead to a zero force [cf. Supplement S1(c)]. Hence, two STs would remain to be considered: Chu's and EL's. But since we cannot differentiate between them inside a body (like e.g. a slab embedded in air [3]), to unify and identify the particular limits of applicability of all ST theories we first consider a sufficiently general situation: that pertaining to the optical momentum transfer of a linearly polarized plane wave, with electric vector: $E_x = e^{i(kz-\omega t)}$, impinging an absorbing sphere embedded in air. In terms of only the EL ST it is indeed possible to calculate the total force from the interior wavefields of such an either moderately absorbing (defined in the main article, section IIIA) or gain generic object embedded in air without adding any hidden quantity [as depicted in Fig. 1s (a), (b) for a dielectric and magneto-dielectric sphere respectively, and also later in Supplement S4 (b), moderate absorption can be considered for these examples with the limit: $(\varepsilon_I / \varepsilon_b) \leq 10$, where $\varepsilon_b$ is the background permittivity]. All wavefields in this work are obtained by Mie calculations without any approximations.

On the other hand, Nelson's ST [4], associated to the conventional Lorentz force, does not appropriately yield such interior force unless hidden quantities are added, [cf. Supplement S1 (b)], as demonstrated in Fig. 1s (b). Similarly, from Fig. 1s (b) we infer that Chu ST with fields inside the sphere only yields the kinetic force, but fails to provide its static part (also cf. [5]). In fact, when the permittivity, $\varepsilon_s$ and permeability $\mu_s$ of a particle have moderate imaginary parts, its interior behaves as a 'kinetic system' actually constituting a *"transparent Einstein box"* [6] [see Eq (1) in the main article and Fig. 1s (a) and (b) in this supplement]. Hence, we herewith argue against the *ad hoc* addition of hidden quantities, and support the EL formulation for 'kinetic systems' [cf. supp. S1 (a-c)] for the time-averaged force:

$$\left\langle \boldsymbol{F}_{\text{Total}}^{\text{Kin system}} \right\rangle = \oint \left\langle \bar{\bar{\boldsymbol{T}}}_{\text{EL}}^{\text{in}} \right\rangle \cdot d\boldsymbol{s} = \int \left\langle \boldsymbol{f}_{\text{EL}}^{\text{in}} \right\rangle dv$$

with the Einstein-Laub ST and force density:

$$\left\langle \bar{\bar{\boldsymbol{T}}}_{\text{EL}}^{\text{in}} \right\rangle = \frac{1}{2} \text{Re} \left[ \boldsymbol{D}_{\text{in}} \boldsymbol{E}_{\text{in}}^* + \boldsymbol{B}_{\text{in}} \boldsymbol{H}_{\text{in}}^* - \frac{1}{2} \left( \mu_0 \boldsymbol{H}_{\text{in}} \cdot \boldsymbol{H}_{\text{in}}^* + \varepsilon_0 \boldsymbol{E}_{\text{in}} \cdot \boldsymbol{E}_{\text{in}}^* \right) \bar{\bar{\boldsymbol{I}}} \right],$$

$$\boldsymbol{f}_{\text{EL}}^{\text{in}}(t) = (\boldsymbol{P} \cdot \nabla) \boldsymbol{E}_{\text{in}} + (\boldsymbol{M} \cdot \nabla) \boldsymbol{H}_{\text{in}} + \frac{\partial \boldsymbol{P}}{\partial t} \times \mu_0 \boldsymbol{H}_{\text{in}} - \frac{\partial \boldsymbol{M}}{\partial t} \times \varepsilon_0 \boldsymbol{E}_{\text{in}}.$$

We remark that although the constitutive parameters of the embedded object are $\varepsilon_s$ and $\mu_s$, in the above equations the permittivity and permeability are those of vacuum: $\varepsilon_0$ and $\mu_0$, respectively. We consider wavefields with time dependence: $\exp(-i\omega t)$.

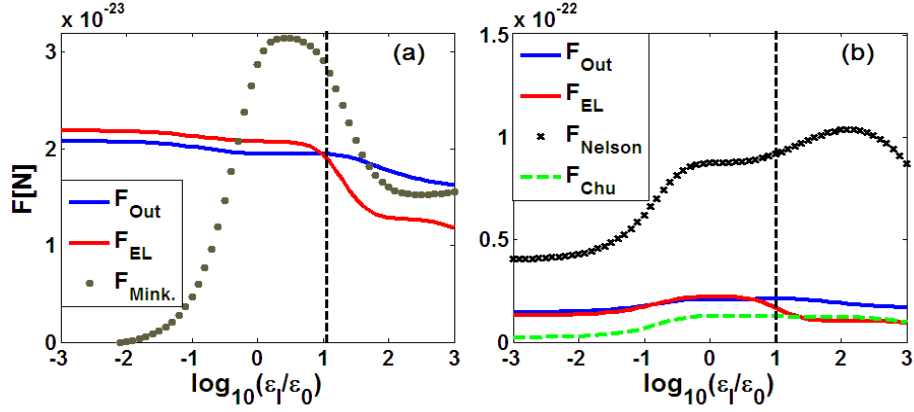

Fig.1s: Time-averaged forces, in Newtons, on a sphere with radius a=1000nm, illuminated by a linearly polarized plane wave: $E_x = e^{i(kz-\omega t)}$. The background is air and $\lambda = 1064$ nm. $F_{out}$ is calculated by the Maxwell stress tensor with outside fields by integrating it on $r = a^+$. $F_{EL}$ is obtained from interior fields by integration of EL ST on the surface: $r = a^-$ [as depicted in Fig. 1 (a) main article]. (a) $\varepsilon_s = 16\varepsilon_0 + i\varepsilon_I$, $\mu_s = \mu_0$ [2]; $F_{out}$ and $F_{EL}$ match well, with only a maximum difference of 5% in the loss region, but Minkowski ST evaluated in $r = a^-$ yields only the conducting force $F_{Mink}$ [2] instead of the total force. (b) $\varepsilon_s = 7\varepsilon_0 + i\varepsilon_I$, $\mu_s = 3\mu_0$. EL ST in $r = a^-$ matches with the outside force $F_{out}$; however, Nelson and Chu STs on $r = a^-$ fail. These results support EL theory inside objects embedded in an air, or vacuum. All calculations contain SI units $\varepsilon_0 = 8.8541878 \times 10^{-12}$ F/m and $\mu_0 = 1.256637 \times 10^{-6}$ H/m.

**S-1: Hidden quantities and previous stress tensors when the body is embedded in air:**

**S-1(a). Chu stress tensor in the interior of an object:**

The Einstein-Laub (EL) force, Chu force and Chu stress tensor inside an object are:

$$f_{EL}(t) = (P \cdot \nabla)E_{in} + (M \cdot \nabla)H_{in} + \frac{\partial P}{\partial t} \times \mu_0 H_{in} - \frac{\partial M}{\partial t} \times \varepsilon_0 E_{in}, \quad (1a)$$

$$f_{Chu}(t) = -(\nabla \cdot P)E_{in} - (\nabla \cdot M)H_{in} + \frac{\partial P}{\partial t} \times \mu_0 H_{in} - \frac{\partial M}{\partial t} \times \varepsilon_0 E_{in}. \quad (1b)$$

$$\overline{\overline{T}}^{in}_{Chu} = \varepsilon_0 E_{in} E_{in} + \mu_0 H_{in} H_{in} - \frac{1}{2}(\mu_o H_{in} \cdot H_{in} + \varepsilon_0 E_{in} \cdot E_{in})\overline{\overline{I}} \quad (1c)$$

The difference between the Chu and EL STs [1] stems from the normal components of the electric and magnetic fields on the body boundaries. Such a difference is only observed in the static part of the force [5]. For instance, a wavefield normally incident on a slab has no normal components of **E** and **H**, thus we cannot differentiate between Chu and EL STs inside this object [3]. However, according to Barnett et al. [5]:*"The problem with $f_c$, (i.e. the Chu force density), appears when we introduce the macroscopic polarization. Averaging **P** over a small volume to get $\langle P \rangle$ is not a problem, but averaging the term $\nabla \cdot P = -Q(t)$ does not generally give $\nabla \cdot \langle P \rangle$. By contrast, the force density $f_d$* (i.e. the EL force inside any dielectric body) *does not contain any spatial derivatives of the microscopic polarization, and so the*

*averaging procedure can be applied safely. An analysis based on the universal form of the Lorentz force suggests, somewhat surprisingly, that the force density is $f_d$ rather than $f_c$*". To solve this problem for general cases, for example Kemp et al. write the static part (strictly on the surface, at r = a or at the interface) of the force in Chu's formulation for a magneto-dielectric as [7]:

$$\langle f_{\text{Static}}^{\text{Chu}} \rangle = \langle f_{\text{Surface}} \rangle = \langle f_e \rangle + \langle f_m \rangle = \frac{1}{2} \text{Re}[\rho_e E_{\text{average}}^* + \rho_m H_{\text{average}}^*]$$

$$= \{\epsilon_o (E_{\text{out}} - E_{\text{in}}) \cdot \hat{n}\} \left(\frac{E_{\text{out}} + E_{\text{in}}}{2}\right)^* + \{\mu_0 (H_{\text{out}} - H_{\text{in}}) \cdot \hat{n}\} \left(\frac{H_{\text{out}} + H_{\text{in}}}{2}\right)^*, \quad (2)$$

where the bound electric and magnetic surface charge densities are [7]: $\rho_e = \{\varepsilon_0 (E_{\text{out}} - E_{\text{in}}) \cdot \hat{n}\}$ and $\rho_m = \{\mu_o (H_{\text{out}} - H_{\text{in}}) \cdot \hat{n}\}$, respectively, $\hat{n}$ being the unit normal to the object surface. $(E_{\text{in}}, H_{\text{in}})$ and $(E_{\text{out}}, H_{\text{out}})$ are the total fields inside and outside the body, respectively, and $E_{\text{average}} = (E_{\text{out}} + E_{\text{in}})/2$ and $H_{\text{average}} = (H_{\text{out}} + H_{\text{in}})/2$. As the term: $[-(\nabla \cdot P)E_{\text{in}} - (\nabla \cdot M)H_{\text{in}}]$ yields almost zero force in $f_{\text{Chu}}$ and does not contribute to the total force inside a continuous object, from Eq (1b) it can be concluded that:

$$\oint \langle \overline{\overline{T}}_{\text{Chu}}^{\text{in}} \rangle \cdot dS \simeq \int \langle f_{\text{kinetic}} \rangle dV$$

$$\text{Where} \quad f_{\text{kinetic}}(t) = \frac{\partial P}{\partial t} \times \mu_0 H - \frac{\partial M}{\partial t} \times \varepsilon_0 E \quad (3a)$$

So, Chu stress tensor shows the behavior that we have named as 'existence domain' in the main article. However, one may arrive the same final force calculated in equation (2) by employing:

$$\langle F_{\text{surface}} \rangle = [\langle \overline{\overline{T}}_{\text{Maxwell}}^{\text{out}} \rangle - \langle \overline{\overline{T}}_{\text{Chu}}^{\text{in}} \rangle] \cdot \hat{n} \quad \text{[Calculated exactly at r = a]} \quad (3b)$$

This happens due to the well known relation: $\langle F \rangle = \int \langle f \rangle dv = \int \langle \nabla \cdot \overline{\overline{T}} \rangle dv$. However, in Eq (3b):

$$\overline{\overline{T}}_{\text{Maxwell}}^{\text{Out}} = \varepsilon_0 E_{\text{out}} E_{\text{out}} + \mu_0 H_{\text{out}} H_{\text{out}} - \frac{1}{2} (\mu_0 H_{\text{out}} \cdot H_{\text{out}} + \varepsilon_0 E_{\text{out}} \cdot E_{\text{out}}) \overline{\overline{I}} \quad (3c)$$

In equation (3c) $\overline{\overline{T}}_{\text{Maxwell}}^{\text{out}}$ is the ST originally derived by Maxwell, accounting for the optical momentum transfer to the body from air or vacuum. For objects embedded in a material background Minkowski ST applies [8], and of course $\overline{\overline{T}}_{\text{Mink}}^{\text{out}}$ is $\overline{\overline{T}}_{\text{Maxwell}}^{\text{out}}$ when this background is air or vacuum. Equation (2) has been formulated (cf. Eq (3b)) considering that Chu's stress tensor yields only the kinetic part of the time averaged total force inside the object, viz.:

$$\oint \langle \overline{\overline{T}}_{\text{Chu}}^{\text{in}} \rangle \cdot dS \simeq \int \langle f_{\text{Kinetic}}^{\text{in}} \rangle dV \quad (= \langle F_{\text{Kinetic}}^{\text{in}} \rangle) \quad (4)$$

In equation (4) the integration is performed inside a body on the surface $r = a^-$, i.e. with interior fields. As a result, considering the equality of Eq (2) and (3b); and the relation: Eq (4), one may finally write:

$$\langle F_{\text{Hidden}}^{\text{Chu}} \rangle = \langle F_{\text{surface}} \rangle = \langle F_{\text{Total}}^{\text{out}} \rangle - \langle F_{\text{Kinetic}}^{\text{in}} \rangle \quad (5)$$

This is ultimately expressed as:

$$\langle \boldsymbol{F}_{total} \rangle = \langle \boldsymbol{F}_{Kinetic}^{In} \rangle + \langle \boldsymbol{F}_{surface} \rangle \tag{6}$$

Nonetheless, equation (6) does not specify whether $\boldsymbol{F}_{total}$ is obtained independently from interior fields, or whether it only constitutes an artificial mechanism to get the total time averaged force by employing outside background fields. Using equations (5) and (6), one then obtains: $\langle \boldsymbol{F}_{total} \rangle$ [in equation (6)] $= \langle \boldsymbol{F}_{Total}^{outside} \rangle$ in equation (5)].

Thus the above calculations mean that Eq (2) constitutes an artificial procedure to obtain the correct total force from exterior fields by *ad hoc* addition of a 'hidden quantity' to the interior kinetic force. Hence, from those equations we cannot derive the total force on a body by only employing only the interior fields (i.e. Chu ST leads only the kinetic force part of the total force). Moreover, to reach a modified Chu formulation for a complex system like a magneto-dielectric object embedded in another magneto-dielectric or dielectric background, Eq (2) has to be reformulated since then the known Chu's stress tensor in Eq (1c) cannot lead to even the correct kinetic part of the force.

### S-1(b). Nelson stress tensor in the interior of an object:

The law associated to Nelson stress tensor [4] is the conventional Lorentz force [4], which inside an object is written as [1, 4]:

$$\boldsymbol{f}_{Nelson/Lorentz}(t) = -(\nabla \cdot \boldsymbol{P})\boldsymbol{E} + \frac{\partial \boldsymbol{P}}{\partial t} \times \boldsymbol{B} + (\nabla \times \boldsymbol{M}) \times \boldsymbol{B} \tag{7}$$

$$\bar{\bar{\boldsymbol{T}}}_{Nelson/Lorentz} = \varepsilon_0 \boldsymbol{E}\boldsymbol{E} + \mu_0^{-1}\boldsymbol{B}\boldsymbol{B} - \frac{1}{2}\left(\mu_0^{-1}\boldsymbol{B} \cdot \boldsymbol{B} + \varepsilon_0 \boldsymbol{E} \cdot \boldsymbol{E}\right)\bar{\bar{\boldsymbol{I}}} \tag{8}$$

The difference between the EL ST and Nelson ST arises from the terms $\boldsymbol{P} \cdot \nabla$ and $\nabla \cdot \boldsymbol{P}$ on taking the surface integral of equation (8) in the interior of a dielectric body. But for pure magnetic or magneto-dielectric media, the difference is due to the 'hidden momentum' of Shockley [8]. Ultimately, the magnetic dipole inside matter is modelled as an Amperian current loop in the Nelson-Lorentz theory. This produces [9] the departure between EL and Nelson formulations. To correctly obtain the time-averaged force inside an embedded object with Lorentz-Nelson formulation, Shockley's hidden momentum inside the particle, as well as the static electric force part given in equation (2), should be remodelled.

### S-1(c). Minkowski ST and Abraham ST in the interior of an object:

Inside an absorbing body embedded in air one writes for the Minkowski ST:

$$\langle \bar{\bar{\boldsymbol{T}}}_{Mink}^{in} \rangle = \frac{1}{2}\text{Re}\left[ \boldsymbol{D}_{in}\boldsymbol{E}_{in}^* + \boldsymbol{B}_{in}\boldsymbol{H}_{in}^* - \frac{1}{2}\left(\boldsymbol{B}_{in} \cdot \boldsymbol{H}_{in}^* + \boldsymbol{D}_{in} \cdot \boldsymbol{E}_{in}^*\right)\bar{\bar{\boldsymbol{I}}} \right] \tag{9}$$

$$\langle \nabla \cdot \bar{\bar{\boldsymbol{T}}}_{Mink}(in) \rangle = \langle \boldsymbol{f} \rangle = \langle \boldsymbol{f}_c \rangle = \frac{1}{2}\text{Re}\left[ \omega\varepsilon_I \boldsymbol{E}_{in} \times \boldsymbol{B}_{in}^* - \omega\mu_I \boldsymbol{H}_{in} \times \boldsymbol{D}_{in}^* \right] \tag{10}$$

Hence, the internal Minkowski ST accounts only for the force felt by the conducting currents due to losses inside the object. In fact, the photon and carrier interaction in the body are considered in the Minkowski photon momentum $\boldsymbol{p}_{Mink}$ [2, 6]. Hence, the hidden quantity for the Minkowskian formulation inside the object is:

$$\left\langle \boldsymbol{F}_{\text{Hidden}}^{\text{Mink.}} \right\rangle = \left\langle \boldsymbol{F}_{b}^{\text{in}} \right\rangle = \left\langle \boldsymbol{F}_{\text{Total}}^{\text{out}} \right\rangle - \left\langle \boldsymbol{F}_{c}^{\text{in}} \right\rangle \tag{11a}$$

$$\left\langle \boldsymbol{F}_{c}^{\text{in}} \right\rangle = \int \left\langle \boldsymbol{f}_{c} \right\rangle dV \tag{11b}$$

In [2] $\boldsymbol{F}_b$ (the bound force) is defined with the formalism that contains $\nabla \cdot \boldsymbol{P}$ and $\nabla \cdot \boldsymbol{M}$, which should be modified by the terms $\boldsymbol{P} \cdot \nabla$ and $\boldsymbol{M} \cdot \nabla$, resulting in a new force density and time-averaged bound force:

$$\boldsymbol{f}_b(\text{new}) = (\boldsymbol{P} \cdot \nabla)\boldsymbol{E}_{\text{in}} + (\boldsymbol{M} \cdot \nabla)\boldsymbol{H}_{\text{in}} + \frac{\partial(\text{Re}[\varepsilon] - \varepsilon_0)\boldsymbol{E}_{\text{in}}}{\partial t} \times \mu_0 \boldsymbol{H}_{\text{in}} - \frac{\partial(\text{Re}[\mu] - \mu_0)\boldsymbol{H}_{\text{in}}}{\partial t} \times \varepsilon_0 \boldsymbol{E}_{\text{in}} \tag{12}$$

$$\left\langle \boldsymbol{F}_{b}^{\text{In}} \right\rangle = \int \left\langle \boldsymbol{f}_{b}(\text{new}) \right\rangle dv \tag{13}$$

It is a well-known fact that the Abraham ST is just the symmetrical version of Minkowski ST. The Abraham ST also leads to the same force $\boldsymbol{F}_c$ as equation (11b), i.e. the conducting force [2] inside the particle. Note that this force, $\boldsymbol{F}_c$ means the 'outside force' [defined as the 'transferred momentum' in the main article section III.B.A] of the electrons embedded inside the object. However, when there are no losses or in other words: nothing is embedded inside the object ($\varepsilon_I = \mu_I = 0$); both the internal Abraham ST and Minkowski ST yield zero forces in the interior of the object [cf.Eq(10)]. This can be well explained based on the 'existence domain' of Minkowski and Abraham ST. So, when $\varepsilon_I = \mu_I = 0$, if we consider internal Abraham or Minkowski ST to yield the total force inside the object, the 'total' internal force should be calculated using the hidden force in Eq (13). Interestingly, Eq (13) is nothing but the EL force for a lossless object embedded in air or vacuum (also cf. Eq (1a)).

Considering all the above mentioned issues mentioned in supplement S0 and S1 (a) –(c), the formulation of our Generalized Einstein-Laub (GEL) equations starts from the EL ST and force to yield the total interior force of an embedded object without hidden quantities. In fact, for more omplex systems such as an arbitrary magneto-dielectric object embedded in a generic heterogeneous magneto-dielectric background, the situation of internal force calculation would become quite complex based on formulations other than the GEL equations proposed in this work.

### S2. A box embedded in a background medium: Failure of the Einstein-Balazs thought experiment

If a rectangular box with refractive index $n_s = c\sqrt{\varepsilon_s \mu_s}$ is embedded in air or vacuum, the thought experiment of Einstein-Balazs can predict the appropriate internal photon momentum, i.e. $\boldsymbol{p}_{\text{Abr}}$ (in), which leads to the appropriate internal stress tensor of Einstein-Laub to obtain the total force on a scatterer excluding hidden quantities. But if the background is non-vacuum, but has a refractive index $n_b = c\sqrt{\varepsilon_b \mu_b}$, does any thought experiment of Einstein-Balazs type predict the appropriate photon momentum inside the box?. The interior dynamics of such an embedded box (i.e. any macroscopic object embedded in another material background) has been defined as the '*generalized Einstein Laub (GEL) system of first kind*' in the main paper. Although the tangential component of the photon momentum is coserved across the boundary, (at the interfaces of two different media, the tangential component of the photon momentum always arises as $\boldsymbol{p}_{\text{Mink}}$, and Snell's law also supports this fact), we investigate in this section the normal component of the photon mmomentum and its involvement in the optical force. Although the transfer of both $p_{Abr}$ (out) and $p_{Mink}$ (out) from the background have been predicted for microscopic objects [10], what happens in the interior of an embedded macroscopic object considering both the transfer of $p_{Abr}$ (out) and $p_{Mink}$ (out) is the ultimate goal of this section.

### S-2 (a). Case-1: Transfer of Abraham's photon momentum from the background

If Abraham's stress tensor is applied to determine the transfer of momentum from the background to an embedded macroscopic object, (i.e. the 'outside force' defined in the main paper), both the Minkowski ST and Abraham ST predict the same total 'outside force'. But no approach that directly measures the transfer of photon momentum from the background to an embedded object [11-14] reveals $p_{Abr}$ (out) as the transferred photon momentum. However, based on the microscopic formula of the 'outside force' [10] (which behaves like the Abraham force law for the macroscopic objects), the transfer of $p_{Abr}$ (out) has been predicted in [10]. If we consider $p_{Abr}$ (out) as the transferred momentum for an embedded macroscopic object, the momentum conservation of an Einstein-Balazs box of length $L$ and mass $M$ is expressed by that of the kinetic momentum [15]:

$$p_{box}^{Kinetic} = M_{box} \cdot V_{box}^{Kinetic} = p_{Photon}(\text{out}) - p_{Photon}(\text{in}), \tag{14}$$

In equation (14) $p$ represents 'momentum'.

$$p_{box}^{Kinetic} = M_{box} \cdot V_{box}^{Kinetic} = m(c/n_b) - m(c/n_s), \tag{15}$$

where $m = E/c^2$ is the mass associated to the photon [15]. Now, we can write:

$$V_{box}^{Kinetic} = \frac{1}{M_{box}} \left[ m\frac{c}{n_b} - m\frac{c}{n_s} \right] = \frac{\Delta z}{\Delta t} \tag{16}$$

$$\Delta t = \frac{L}{c/n_s}, \quad E = mc^2 = \hbar\omega$$

$$\Delta z = V_{box}^{Kinetic} \cdot \Delta t = \frac{\hbar \omega L}{Mc^2}(n_s n_b^{-1} - 1), \tag{17}$$

$$p_{box}^{kinetic} = M_{box} \cdot \frac{\Delta z}{\Delta t} = \frac{\hbar \omega}{c}(n_b^{-1} - n_s^{-1}) \tag{18}$$

Thus from Eq (14) and considering $p_{Photon}(\text{out}) = p_{Abr}(\text{out})$, the internal photon momentum is revealed as:

$$p_{photon}(\text{in}) = p_{Abr}(\text{in}) \tag{19}$$

Thus, if $p_{Abr}$ (out) is considered as the transferred momentum to an embedded macroscopic object, the internal momentum of the embedded box is revealed as $p_{Abr}$ (in). But according to Eq (18), if $n_b > n_s$, the embedded box will be pulled due to simple plane wave illumination, which has not been observed in any experiment. Moreover, any internal ST, that supports $p_{Abr}$ (in) (i.e. the Einstein-Laub and Chu ST), does not lead to the internal correct time avergaed total force on an embedded macroscopic object. As a result, Eq (17) may not represent the actual displacement of the box, and hence Eq (18) may not be a valid result.

### S-2 (b). Case-2: Transfer of Minkowski photon momentum from the background

If it is considered that the transferred photon momentum to an embedded macroscopic object (i.e. an embedded Einstein-Balazs box) from the background is the one of Minkowski [11-14], from Eq (14) we can write the total momentum conservation as:

$$\boldsymbol{p}_{box} = M_{box} \cdot \boldsymbol{V}_{box} = [m(c/n_b) + \boldsymbol{p}_{med}(out)] - m(c/n_s), \quad (20)$$

Note that to differentiate the mechanical momentum and the velocity of the embedded Einstein-Balazs box, the notations are different in Eq (14) and in Eq (20). However, in Eq (20), we have used $[m(c/n_b) + \boldsymbol{p}_{med}(out)] = \boldsymbol{p}_{Mink}(out) = \boldsymbol{p}_{Photon}(out)$ according to our note [46] and [47] in the main paper. So, from Eq (20), we can write:

$$\boldsymbol{V}_{box} = \frac{1}{M_{box}} \left[ m\frac{c}{n_b} - m\frac{c}{n_s} \right] + \frac{1}{M_{box}} \boldsymbol{p}_{med}(out) \quad (21)$$

Now, using the definition: $\boldsymbol{p}_{med}(out) = (n_b - n_b^{-1})\hbar\omega/c$ in Eq (21) (cf. the main paper) and $E = mc^2 = \hbar\omega$, we can write:

$$\boldsymbol{p}_{Box}(in) = M_{box} \cdot \boldsymbol{V}_{box} = (n_b n_s - 1)(\hbar\omega/n_s c) \quad (22)$$

And employing Eq (22) and $\boldsymbol{p}_{Mink}(out) = \boldsymbol{p}_{Photon}(out) = n_b(\hbar\omega/c)$ in Eq (14), the internal photon momentum is revealed as:

$$\boldsymbol{p}_{photon}(in) = \boldsymbol{p}_{Abr}(in) \quad (23)$$

According to Eq (22), the embedded Einstein-Balazs box does not experience the pulling force from the incident light, and the velocity of photons inside the box is $c/n_s$. Although this result seems more accurate than the previous case in Supplement S2 (a), no stress tensor nor force density formula related with the Abraham photon momentum leads to the correct interior force on an embedded object. This is the reason why we need a more rigorous formulation to attain any final conclusion regarding the internal photon momentum of an embedded Einstein-Balazs box, or in more complex situations. This is done in the next section. Another important question is why $\boldsymbol{p}_{Abr}(in)$ appears as the fixed photon momentum inside the embedded Einstein-Balazs box irrespective of the transferred photon momentum from the background (cf. also our note [16] below). This is also addressed in the next section.

### S3. The interior photon momentum of a '*generalized Einstein Laub (GEL) system*' and the derivation of the Generalized Einstein-Laub (GEL) force:

In Supplement S2 we have shown that the Einstein-Balaz's type thought experiment may not lead to the appropriate internal photon momentum of an embedded object (that is defined as '*generalized Einstein Laub* system of first kind' in the main paper). On the other hand, the arguments in favor of non-relativistic Doppler-shift effects [12, 15] always lead to the transfer of photon momentum from the background to an embedded object, rather than to the internal photon momentum of any embedded object. In this section, rather than following the thought experiment, we argue on the internal photon momentum of a '*generalized Einstein Laub* system' (defined at the beginning in section III. C in the main paper) based on the more rigorous approach of the linear momentum conservation equation: $\frac{\partial \boldsymbol{p}_{\text{Total}}}{\partial t} = \nabla \cdot \bar{\bar{\text{T}}} = \boldsymbol{f} + \frac{\partial}{\partial t}\boldsymbol{G}$.

Here $\boldsymbol{p}_{\text{Total}}$ is the total momentum of the system. Note that the '*generalized Einstein Laub* system' covers several cases where the internal photon momentum of a generic object is not the one of Abraham. Those cases are: (i) generalized *Einstein Laub* system of *first kind*: the internal dynamics of an object embedded in another material background, (ii) *generalized Einstein Laub* system of *second kind*: the internal dynamics of a highly absorbing object embedded in simple air or vacuum background, (iii) *generalized Einstein Laub* system of *third kind*: the internal dynamics of a chiral object embedded in another material background. Note that a 'kinetic system' is only a special case of a '*generalized Einstein Laub* system'.

### S3 (a). Transitional Generalized Einstein-Laub equations

Since we are not dealing with a 'kinetic system', inside the embedded Einstein-Balazs box the EL equations should be modified. To do so, we first write the relationship between the internal Minkowski ST and the EL ST in terms of the internal fields $\boldsymbol{E}_{\text{in}}$, $\boldsymbol{H}_{\text{in}}$, polarization $\boldsymbol{P}_{\text{in}}$ and magnetization $\boldsymbol{M}_{\text{in}}$:

$$\bar{\bar{T}}_{\text{EL}}^{\text{in}} = \bar{\bar{T}}_{\text{Min}}^{\text{in}} + \frac{1}{2}\left(\boldsymbol{M}_{\text{in}} \cdot \boldsymbol{H}_{\text{in}} + \boldsymbol{P}_{\text{in}} \cdot \boldsymbol{E}_{\text{in}}\right)\bar{\bar{I}} \tag{24a}$$

The field momentum density being that of Abraham, viz.:

$$\boldsymbol{G}_{\text{EL}}(\text{in}) = \boldsymbol{G}_{\text{Abr}}(\text{in}) = [\boldsymbol{D}_{\text{in}} - \boldsymbol{P}_{\text{in}}] \times [\boldsymbol{B}_{\text{in}} - \boldsymbol{M}_{\text{in}}] \tag{24b}$$

When the embedding background is a magneto-dielectric medium instead of air, the conventional vectors $\boldsymbol{P}$ and $\boldsymbol{M}$ inside the particle should be replaced by what we call *effective* polarization $\boldsymbol{P}_{\text{Eff}}$ and magnetization $\boldsymbol{M}_{\text{Eff}}$, respectively (see Section III.C.a of the main paper). Thus, we put forward a more universal form of the EL ST, equation (24a), by expressing it inside any material object in terms of the *effective* vectors $\boldsymbol{P}_{\text{Eff}}$ and $\boldsymbol{M}_{\text{Eff}}$, (cf. Eqs (7a) and (7b) in the main paper). This is our generalized Einstein-Laub (GEL) ST:

$$\bar{\bar{T}}_{\text{GEL}}^{\text{in}} = \bar{\bar{T}}_{\text{Minkowski}}^{\text{in}} + \frac{1}{2}\left(\boldsymbol{M}_{\text{Eff}} \cdot \boldsymbol{H}_{\text{in}} + \boldsymbol{P}_{\text{Eff}} \cdot \boldsymbol{E}_{\text{in}}\right)\bar{\bar{I}} \tag{25a}$$

Which can be written for any generic object as:

$$\bar{\bar{T}}_{\text{GEL}} = \boldsymbol{D}_{\text{in}}\boldsymbol{E}_{\text{in}} + \boldsymbol{B}_{\text{in}}\boldsymbol{H}_{\text{in}} - \frac{1}{2}\left(\mu_{\text{Eff}}\boldsymbol{H}_{\text{in}} \cdot \boldsymbol{H}_{\text{in}} + \varepsilon_{\text{Eff}}\boldsymbol{E}_{\text{in}} \cdot \boldsymbol{E}_{\text{in}}\right)\bar{\bar{I}} \tag{25b}$$

We now address the continuity of the instantaneous momentum. We obtain from equation (25b):

$$\nabla \cdot \bar{\bar{T}}_{\text{GEL}} = (\boldsymbol{P}_{\text{Eff}} \cdot \nabla)\boldsymbol{E} + (\boldsymbol{M}_{\text{Eff}} \cdot \nabla)\boldsymbol{H} - \varepsilon_{\text{Eff}}\boldsymbol{E} \times (\nabla \times \boldsymbol{E}) - \mu_{\text{Eff}}\boldsymbol{H} \times (\nabla \times \boldsymbol{H}) \tag{26}$$

Similar to equation (24b), we write the GEL momentum density inside an object:

$$\boldsymbol{G}_{\text{GEL}}^{\text{Transition}}(\text{in}) = [\boldsymbol{D}_{\text{in}} - \boldsymbol{P}_{\text{Eff}}] \times [\boldsymbol{B}_{\text{in}} - \boldsymbol{M}_{\text{Eff}}] \tag{27a}$$

However the non-mechanical momentum density that appears in Eq (27a) is not the definitive form of this quantity. For a '*generalized Einstein Laub* system', defined previously, Eq (27a) is applicable inside any generic object to calculate a transitional part of the time-averaged total final force. For this reason, we have introduced the term 'transition'. The only significance of Eq (27a) is that if the background is air or vacuum, this momentum density exactly matches with the Abraham momentum density inside any scatterer with no or moderate absorption. However, in order to match the continuity equation for an embedded body, we get from Eq (27a) and from Eqs (7a) and (7b) of the main paper that:

$$G_{GEL}^{Transition}(in) = \varepsilon_{Eff}\mu_{Eff}\left(E_{in} \times H_{in}\right) = \left[G_{Material}^{Transition} + G_{Electromag}\right] = [\varepsilon_{Eff}\mu_{Eff} - 1/c^2]\left(E_{in} \times H_{in}\right) + \left(E_{in} \times H_{in}\right)/c^2 \quad (27b)$$

To investigate the physical meaning of equation (27b), we have splitted it into two parts. In terms of the photon momentum, equation (27b) can be written as:

$$p_{Non-Mech.}^{Transition}(in) = p_{Material}^{Transition}(in) + p_{Abr}(in) \quad (27c)$$

Here $p_{Non-Mech.}^{Transition}(in) = \int G_{GEL}^{Transition}(in)\,dv$ and $p_{Abr} = \int G_{Abr}\,dv$. $p_{Material}^{Transition}(in)$ is defined as the *material induced momentum*. The transitional mathematical expression (but not the effective mathemtical expression) of $p_{Material}^{Transition}(in)$ in Eq (27c) can be written as: $p_{Material}^{Transition}(in) = \int G_{Material}^{Transition}(in)\,dv = \left(c^2\varepsilon_{Eff}\mu_{Eff} - 1\right)p_{Abr}(in)$. The non-mechanical momentum $p_{Non-Mech.}(in)$ of the 'generalized *Einstein Laub* system' in Eq (27c) is neither the one of Abraham nor of Minkowski. The importance of Eq (27c) is seen as we show that the variation of the magnitude of the non-mechanical momentum occurs inside any generic object due to the variation of the magnitude of the material induced momentum part in Eq (27c), though the mathematical form of $p_{Abr}(in)$ should always remain the smae. This 'magnitude variation' is expressed as

$$p_{Material}(in) = \int G_{Material}(in)\,dv = \left(c^2\varepsilon_{Eff}\mu_{Eff} - 1\right)p_{Abr}(in) + \text{Additional term} \quad (27d)$$

The 'Additional term' in Eq (27d) is characterized in the next section, [Eq (30c)]. According to Eq (27c), due to the variation of the 'material induced momentum', the mechanical momentum of a generic scatterer may also vary. The question is how to account for such variations of the mechanical momenta of the generic scatterer in the ultimate force law. This is next addressed based on the time-average force law [17] rather than on the time-varying optical force equation.

### S3 (b). Effective Generalized Einstein-Laub Equations

To determine the final form of the total force $\langle f_{GEL}\rangle = \langle f_{GEL}^{Transition} + \text{Correction term}\rangle$, at first we have to determine that transitional part, $f_{GEL}^{Transition}$, of the total force from Eq (27b). To do so, we address Maxwell's equations inside any generic object:

$$\nabla \times E_{in} = -\frac{\partial B_{Eff}}{\partial t}, \quad \nabla \times H_{in} = \frac{\partial D_{Eff}}{\partial t}, \quad \nabla \cdot D_{Eff} = 0, \quad \nabla \cdot B_{Eff} = 0 \quad (28)$$

Using equations (7a), (7b) of the main paper and (26), (27b) and (28) of this Supplement, we arrive at the following transitional GEL force density:

$$f_{GEL}^{Transition} = (P_{Eff}\cdot\nabla)E_{in} + (M_{Eff}\cdot\nabla)H_{in} + \frac{\partial P_{Eff}}{\partial t}\times\mu_{Eff}H_{in} - \frac{\partial M_{Eff}}{\partial t}\times\varepsilon_{Eff}E_{in} \quad (29)$$

For the '*generalized Einstein Laub* system', Eq (29) represents only the transitional part of the total force. We verify in Supplement S4 (a) that for a magneto-dielectric slab embedded in another magneto-dielectric, Eq (29) does not lead to a correct total internal force, although the ST equation (25b) remains valid for this case. Even for a dieletric object embedded in another dielectric [cf. curve C in Figs.1 and 2 in [18] where a special form of our forthcoming Eq (30a) predicts the correct result with $\varepsilon_{\text{Eff}} = \varepsilon_b$], or on the interface between two dielectrics [11], [where our Eq (25b) predicts the correct result with $\varepsilon_{\text{Eff}} = \varepsilon_b$ and $\mu_{\text{Eff}} = \mu_b$], Eq (29) does not lead to the correct result. The only significance of Eq (29) is that if the background is air or vacuum, this force law exactly matches with the Einstein-Laub force inside any scatterer with no or moderate absorption.

The physical effect of the variation of mathematical expressions (i.e. magnitudes) of both the mechanical and non-mechanical momenta of a whole system, can be accounted in the force law in this way: due to the variation of $p_{\text{Material}}(\text{in})$, the conventional vectors $P$ and $M$ inside the object are modified to $P_{\text{Eff}}$ and $M_{\text{Eff}}$. Henceforth, $P_{\text{Eff}}$ and $M_{\text{Eff}}$ act as effective, or functional, sources and induce the currents $J^{\text{Electric}} = \partial P_{\text{Eff}}/\partial t$ and $J^{\text{Magnetic}} = M_{\text{Eff}}/\partial t$, respectively. These $P_{\text{Eff}}$ and $M_{\text{Eff}}$ start to interact with the newly induced currents $J^{\text{Magnetic}}$ and $J^{\text{Electric}}$ respectively inside any generic object. Hence, we introduce an additive Correction term: $\partial(P_{\text{Eff}} \times M_{\text{Eff}})/\partial t$ as written in Eq (29). This term should be substracted from the $\mathbf{G}_{\text{GEL}}^{\text{Transition}}(\text{in})$ defined in Eq (27a) to maintain the rate of change of total momentum fixed, (see the argument in note [19] below).

The 'additional part' of the total force due to the Correction term $\partial(P_{\text{Eff}} \times M_{\text{Eff}})/\partial t$ can be considered as an ultimate effect of the non-mechanical momentum that turns into a 'mechanical momentum' inside the embedded Einstein-Balazs box. Hence the final time-averaged force acts on the object by interaction between $J^{\text{Electric}}$ and $B$ [17, 20], and between $J^{\text{Magnetic}}$ and $D$ [17]. Thus, introducing equations (7a) and (7b) of the main paper into equation (29) above, taking time-averages, and taking real parts, we finally obtain the GEL time-averaged force $\langle f_{\text{GEL}} \rangle$ (cf. Eq (10a) of the main paper):

$$\langle f_{\text{GEL}} \rangle = \frac{1}{2}\text{Re}\left[ \left(P_{\text{Eff}} \cdot \nabla\right)E_{\text{in}}^* + \left(M_{\text{Eff}} \cdot \nabla\right)H_{\text{in}}^* - i\omega\left(P_{\text{Eff}} \times B_{\text{in}}^*\right) + i\omega\left(M_{\text{Eff}} \times D_{\text{in}}^*\right) \right] \quad (30a)$$

where $B_{\text{in}}$ and $D_{\text{in}}$ should be written as $\mu_s H_{\text{in}}$ and $\varepsilon_s E_{\text{in}}$, respectively, rather than splitting them into field and material parts. Hence, the goal of our formulation without considering the background fields is completed. Also note that the ST in Eq (25b), which is connected with the total momentum $\left[\frac{\partial p_{\text{Total}}}{\partial t} = \nabla \cdot \overline{\overline{\text{T}}} = f + \frac{\partial}{\partial t}G\right]$, remains in the same form of Eq (25b) at any instant. So, considering the loss of the $\partial(P_{\text{Eff}} \times M_{\text{Eff}})/\partial t$ term from the non-mechanical momentum density, the effective, or functional, non-mechanical momentum density should be expressed from Eq (27b) as:

$$G_{\text{GEL}}(\text{in}) = \left[G_{\text{Material}} + G_{\text{Electromag}}\right] = [\varepsilon_{\text{Eff}}\mu_{\text{Eff}}\left(E_{\text{in}} \times H_{\text{in}}\right) - (P_{\text{Eff}} \times M_{\text{Eff}})] = (\varepsilon_{\text{Eff}}\mu_s + \varepsilon_s\mu_{\text{Eff}} - \varepsilon_s\mu_s)(E_{\text{in}} \times H_{\text{in}}). \quad (30b)$$

The material and electromagnetic part of $G_{\text{GEL}}(\text{in})$ can be recognized respectively as: $G_{\text{Material}} = (\varepsilon_{\text{Eff}}\mu_s - \varepsilon_s\mu_{\text{Eff}} - \varepsilon_s\mu_s - 1/c^2)(E_{\text{in}} \times H_{\text{in}})$ and $G_{\text{Electromag}} = G_{\text{Abr}} = \left(E_{\text{in}} \times H_{\text{in}}\right)/c^2$. The time average of this non-mechanical momentum density $G_{GEL}(in)$ is zero, as it should be. However, the most interesting fact is that the time-averaged force equation (30a) is applicable irrespective of the non-mechanical momentum density expression (if the appropriate $P_{Eff}$ and $M_{Eff}$ are recognized for different systems). Now, Eq (27c) should be expressed as (cf. also Eq (27d):

$$p_{Non-Mech.}(in) = p_{Material}(in) + p_{Abr}(in) \tag{30c}$$

$p_{Material}(in)$ is the effective material induced momentum which, using Eq (30b), should finally be expressed as: $p_{Material}(in) = \int [(\varepsilon_{Eff}\mu_{Eff} - 1/c^2)(E_{in} \times H_{in}) - (P_{Eff} \times M_{Eff})]dv$.

So, in the '*generalized Einstein Laub* system', though the pure electromagnetic part of that non-mechanical photon momentum always maintains a fixed mathematical expression for $p_{Abr}(in)$, the expression (i.e. magnitude) of the material induced momentum part [which is attributed to total photon momentum, cf. Eq (27c) and (30c)] of the non-mechanical momentum varies due to the different appropriate values of $\varepsilon_{Eff}$ and $\mu_{Eff}$ in different systems such as: GEL system of 1$^{st}$, 2$^{nd}$ and 3$^{rd}$ kind (defined previously); and also kinetic system. These different values have been made clear in the main paper and also in Supplements S4 (a)-(d). In consequence, the mechanical momentum of a generic scatterer also varies for different systems and hence the governing equation of the total momentum conservation should be expressed by a generalized symbolic equation as:

$$p_{Total} = p_{Mechanical}^{med}(in) + p_{Non-Mech.}(in). \tag{30d}$$

Where $p_{Mechanical}^{med}(in)$ and $p_{Non-Mech.}(in)$ are respectively the mechanical momentum of the generic scatterer and the non-mechanical photon momentum inside the scatterer. According to Eq (30d), although the total momentum of the whole system is constant, the electromagnetic and mechanical momenta may take different values even for a single system [19] (and obviously different values for different systems, as $\varepsilon_{Eff}$ and $\mu_{Eff}$ take on different values for different systems). An example of such cases is the internal dynamics of the embedded Einstein-Balazs box (*generalized Einstein Laub* system of first kind), which is dicussed in short in our note [21] below and in Supplement S4 (a), (b) and (c). On the other hand, the '*generalized Einstein Laub* system of second kind' is discussed in supplement S4 (d).

It should be noticed that in Ref. [22] below it is stated: "*We conclude by noting that a number of further momenta have been proposed, with the aim of resolving the Abraham-Minkowski dilemma [2]. By demonstrating the need for two "correct" momenta and associating these, unambiguously, with the Abraham and Minkowski forms, we may hope that we have also removed the need for further rival forms.*"

But according to our analysis, though the measurment of the transferred photon momentum can be considered to yield the one of Minkowski for different cases, the determination of the total internal force (or force distribution) in different systems may lead to different forms of non-mechanical momenta other than the one of Abraham. This is why we have introduced the concept: '*generalized Einstein Laub system*'; and our generalized Einstein-Laub equations should lead to the appropriate total interior force if the appropriate $P_{Eff}$ and $M_{Eff}$ are recognized in each case.

## S-4. Rigorous verifications supporting our Generalized Einstein-Laub theory and establishing the limit of 'moderate loss':

In Supplement S0 we have stated that if and only if one measures the total internal force (or force distribution) on an object with no or moderate absorption, embedded in air or vacuum, Abraham photon momentum emerges as the appropriate electromagnetic momentum. However, there are several other cases where such measurements do not lead to this momentum. In this section, we show several such examples, which support our generalized Einstein-Laub theory and the concept of generalized *Einstein Laub* system proposed in supplement S-3 b. The g*eneralized Einstein Laub* system of first kind (see Fig. 1(c) of the main paper) is dicussed in sub-sections S-4 (a), (b) and (c). The '*generalized Einstein Laub* system of second kind' is addressed in sub-section S-4 d. We stress here that the Generalized Einstein-Laub equations are named as the 'first kind generalized Einstein-Laub ('GEL1') equations' for '*generalized Einstein Laub s*ystems of first kind' in the main paper. On the other hand, the GEL equations are named as the 'second kind generalized Einstein-Laub ('GEL2') equations' for '*generalized Einstein Laub s*ystems of second kind'

**S-4(a). Slab embedded in a dielectric or magneto-dielectric background:**

As an illustration, for a lossless magneto-dielectric slab of sides at *z=0* and *z=d,* embedded in a magneto-dielectric medium, illuminated at normal incidence by a linearly polarized plane wave propagating along the *z* direction with time harmonic dependence: $e^{i(kz-\omega t)}$, the time-averaged force obtained from either fields inside the slab via our GEL1 ST Eq (14) in the main paper, or our time-averaged force (cf. Eq (17) in the main paper), is the same, [with $\varepsilon_{\text{Eff}} = \varepsilon_b$ and $\mu_{\text{Eff}} = \mu_b$ in Eqs (25b) and (30a)]:

$$\left\langle \boldsymbol{F}_{\text{Total}}^{\text{Embedded}}(\text{in}) \right\rangle = \frac{1}{2} \frac{E_0^2}{\mu_s \left(\frac{\mu_b \varepsilon_s}{\mu_s \varepsilon_b} - 1\right)} \left[ (\varepsilon_s - \varepsilon_b) \mu_s - (\mu_s - \mu_b) \varepsilon_s \right] \left\{ 1 + |R|^2 - |T|^2 \right\}, \tag{31a}$$

Where $\left\langle \boldsymbol{F}_{\text{Total}}^{\text{Embedded}}(\text{in}) \right\rangle$ is the force in per unit area. Here:

$$|R|^2 = \frac{\left(\frac{\mu_b \epsilon_s}{\mu_s \epsilon_b} - 1\right)^2 [\sin(k_s d)]^2}{4\left(\frac{\mu_b \epsilon_s}{\mu_s \epsilon_b}\right)[\cos(k_s d)]^2 + \left(1 + \frac{\mu_b \epsilon_s}{\mu_s \epsilon_b}\right)^2 [\sin(k_s d)]^2} \tag{31b}$$

$$\boldsymbol{E}_x = \begin{cases} \left(e^{jk_b z} + R \cdot e^{-jk_b z}\right) \boldsymbol{E}_0 & z > 0 \\ \left(a e^{jk_s z} + b e^{-jk_s z}\right) \boldsymbol{E}_0 & 0 < z < d \\ T \cdot e^{jk_b z} \boldsymbol{E}_0 & z < d \end{cases}$$

$$|T|^2 + |R|^2 = 1$$

$\boldsymbol{R}$ and $\boldsymbol{T}$ denote the reflection and transmission coefficients of the slab, while *a* and *b* are constants determined from the boundary conditions. The most important observation here is that if we calculate the time-averaged force using equation (29) of this Supplement, it does not match with the GEL1 ST calculation, Eq (14) in the main paper. A similar but not identical observation has been recently reported by Kemp et al. [17] to distinguish between time-dependent and the time-averaged forces. It appears that indeed the time-averaged force acts on the particle by interaction between $\boldsymbol{J}^{\text{Electric}}$ and $\boldsymbol{B}$,[17,20], and between $\boldsymbol{J}^{\text{Magnetic}}$ and $\boldsymbol{D}$ [17], respectively.

Another important issue is to verify the agreement of the total interior GEL1 force with that from the external Minkowski ST. To simplify, let us consider $\mu_s = \mu_b = \mu_0$ in the above calculations. In [23], based on the Minkowski ST approach, it has been shown that the total time-averaged external force for a lossless slab is $\langle F_{\text{Total}}^{\text{Embedded}}(\text{out}) \rangle = \frac{1}{2}\varepsilon_b E_0^2 \left(1 + |R|^2 - |T|^2\right)$, which exactly matches with our internal GEL1 ST and time-averaged force result given in equation (31a). However, this force can also be calculated on using a direct approach based on the external photon momentum. For example, we can consider a beam normally incident on the dielectric slab embedded in another dielectric. It has a photon flux $N_i = \langle S \rangle / \hbar\omega$, where the time-averaged Poynting vector is: $\langle S \rangle = \frac{1}{2}\sqrt{\varepsilon_b/\mu_0}|E_0|^2$; being the momentum flux [24]: $\tau_i = N_i \hbar k_i$. The reflected beam will have a momentum $\tau_r = N_r \hbar k_r$, where $N_r = |R|^2 N_i$. The transmitted (emitted) beam with $N_t$ photons has a momentum flux $\tau_t = N_t \hbar k_t$, where $N_t = |T|^2 N_i$. The total force per unit area applied to the dielectric then is:

$$\langle F_{\text{Total}}^{\text{Embedded}}(\text{out}) \rangle = \tau_i + \tau_r - \tau_t = N_i \hbar k_i + N_r \hbar k_r - N_t \hbar k_t \tag{31c}$$

Hence, using $N_i$, the result exactly matches with equation (31a). Equation (31c) is the fundamental equation of the interfacial tractor beam (ITB) concept. If the background of the input interface of the slab is air and that of the output interface is water, then only within the Minkowskian approach one will have:

$$\tau_t > \tau_i + \tau_r \tag{32}$$

which according to equation (31c) will cause an optical pulling on the slab. A recent interfacial tractor beam experiment supports this fact [11,14]. For a simple bi-background case like [11] and [14], the $|T|^2$ value for the bi-background can be lower than the case of the $|T|^2$ value for the single air background. Hence the only way to consider the pulling effect is attributing $p_{\text{Mink}}$ (out) in Eq (31c) [instead of $p_{\text{Abr}}$ (out)] for the momentum transfer from the background. It is important to note that the conventional Einstein-Balazs' thought experiment cannot explain ITB effect [14]. By contrast, the latter is succesfully addressed by our generalized EL theory in supplementt S3, [cf. also main paper and Supplement S4 (c)].

**S-4(b). Gain particle embedded in air or in a magneto-dielectric background:**

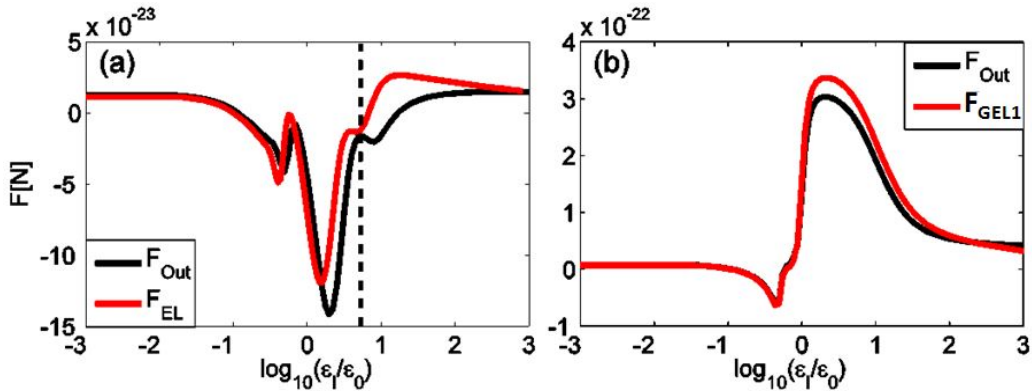

Figure 2s: Illustration of optical momentum transfer from an incident plane wave with electric vector $E_x = e^{i(kz-\omega t)}$ to a gain sphere with $a = 1000$ nm, $\varepsilon_s = 15\varepsilon_0 - i\varepsilon_I$, $\mu_s = \mu_0$; $\lambda = 1064$ nm. (a) Air background. Time-averaged forces: $F_{\text{out}}$ in Newtons, calculated using the Maxwell ST, equation (3b), on the sphere

surface $r=a^+$ using exterior fields, and $F_{in}$ based on the EL ST on the surface $r = a^-$ using interior fields. (b) Magneto-dielectric background: $\varepsilon_b = 4\varepsilon_0$, $\mu_b = 2\mu_0$. Time-averaged forces: $F_{out}$ calculated using Minkowski ST on the surface $r = a^+$ with exterior fields, and $F_{in}$ based on GEL1 ST at the surface $r = a^-$ using interior fields.

Fig. 2s shows that the time averaged force obtained from the EL ST with fields inside a gain particle embedded in air, matches with that calculated from the outside fields with the Maxwell ST, Eq.(3b) only upto the moderate loss limit: $(\varepsilon_I / \varepsilon_b) \leq 10$, which matches with the previously defined limit shown in Fig. 1s (a), (b). On the other hand, if the background is magneto-dielectric still the time-averaged force from the GEL1 ST with the inner fields matches with that obtained via the Minkowski ST with the background wavefields. It is a little bit surprising that even after crossing the approximate moderate loss limit: $(\varepsilon_I / \varepsilon_b) \leq 10$, the 'outside force' by Minkowski ST matches with the internal force by GEL1 ST.

### S-4(c). Object embedded in a heterogeneous background:

The success of the GEL1 formulation and the generalized Einstein-Laub theory are also seen in connection with Fig. 3(b) of the main paper. The object under consideration is a magneto-dielectric infinite cylinder with radius $a=2000$ nm. 25% of the object is submerged in each of the four magneto-dielectric backgrounds (cf. Fig. 3s). Thus the background is now heterogeneous. The GEL1 ST should be employed in regions 1, 2, 3 and 4 inside the embedded object (by employing only interior fields), but with four different permittivities and permeabilities of the local, or inmediate, backgrounds corresponding to those four different interior regions (cf. Eq (14) in the main paper). These complex situations have been successfully handled by our 2D full wave simulation, and the results have been shown in Fig. 3(b) of the main paper. It is important to note that the interior force dynamics of the embedded scatterer in such a heterogeneous medium is indeed not only dependent on the refractive index of the first and the last medium, (which is the fourth medium in Fig. 3s). Rather all the immediate background media are contributing, which has effects in the interior dynamics, i.e. in the 'force felt' by the scatterer.

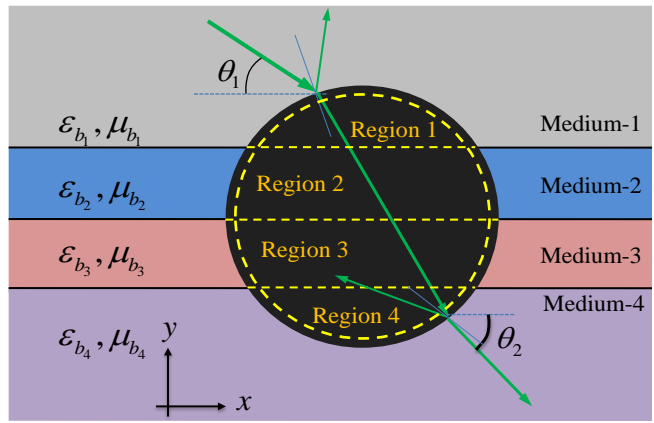

Figure 3s: Schematics of the force exerted on a magneto-dielectric infinite cylinder when light (a plane wave) is scattered in four different magneto-dielectric backgground media. $\theta_1$ and $\theta_2$ stand for the angles of incidence and refraction with the $x$-axis. The incident plane wave is linearly polarized and its angle of incidence $\theta_1$ varies from 0 to $\pi$, considering it positive with the negative $x$-axis, (cf. the results of Fig. 3(b) in the main paper).

Interestingly, while the momentum $p_{Abr}$ (out) with exterior fields leads to a pushing scattering force,

$p_{\text{Mink}}$ (out) (which is the 'field' plus 'mechanical' momentum) predicts a negative scattering force with only 8% inaccuracy with respect to the recent experiments on a bi-background reported in [14]. But an inherent problem of the set-up in [11] or [14] is that the first medium is air and the last medium is water. Hence the effect of $\boldsymbol{P}_{med}$(out) is noticeable only in the last medium, which should always lead to a pulling force [11]. But the idea presented here for the heterogeneous background, i.e. the effect of $\boldsymbol{P}_{\text{med}}$(out), is noticeable for all the four backgrounds. Even if the shape of the scatterer were not a sphere (not shown in this paper), still for the inside force calculation on using the GEL1 ST, the parameters of all the four backgrounds should be considered. As a result, some situations may arise where a ray tracing method [11,14] that may lead to incorrect results while force calculations based on the time-averaged STs (i.e., Minkowski ST from outside and GEL1 ST from inside) remain a general procedure to predict the correct pushing, or pulling, effect on a generic object, (i.e.either a Rayleigh, dipole, Mie, or more complex than Mie, object).

Undoubtedly, the internal EL ST or any other force that supports $p_{\text{Abr}}$ (in) does not account for the effect of the four backgrounds and they do not lead to the total internal force of an object embedded in a heterogeneous background. As a result, the ST, force and non-mechanical momentum density presented in supplement S3 (b) (e.g.the generalized Einstein-Laub theory) should be considered as more general formulations than the Einstein-Laub ST, force and Abraham momentum density.In addition, Eqs. (31c) and (32) also explain why $p_{\text{Abr}}$ (out) should not be employed in the backgrounds of Fig. 3s for analyzing the momentum transfer from the backgound into the object, [see also our discussion in the last part of Supplement S4 (b)]. So, our final conclusion: two fully different dynamics are involved for transferred momentum (i.e. Minkowski's equations) and delivered momentum (i.e. GEL1 equations) respectively, which can be explained successfully by our proposal of 'existence domain' of STs and photon momenta.

### S-4(d). G*eneralized Einstein-Laub* **system of second kind: Highly absorbing objects embedded in air**

[Before the beginning of this sub-section, it should be noted that we have verified our forthcoming conclusions regarding highly absorbing objects (embedded in air) for several other cases not shown here for brevity. The conclusions drawn here remain valid for all those instances, too. Fig. 4s is only a model example among them to introduce the concepts connected with Eqs (22a)-(22c) in the main paper.]

In Fig-1s (a) and (b) of Supplement S0, we see that for a dielectric or magneto-dielectric object, the EL force associated with $p_{\text{Abr}}(\text{in})$ matches with the force found by the exterior Maxwell ST only up to moderate losses (approximately: $(\varepsilon_I / \varepsilon_b) \leq 10$; here $\varepsilon_b = \varepsilon_0$). But in Fig.4s we have considered force (both the external and internal) on a highly absorbing magneto-dielectric object [i.e. that crosses beyond the moderate loss limit (i.e. with $(\varepsilon_I / \varepsilon_b) > 10$)] embedded in air or vacuum. Its internal force is calculated based on the GEL2 ST, Eq (23), of the main paper. In Fig.4s we observe that only when the internal abosrption of an extremely lossy scatterer (i.e. approximately with $(\varepsilon_I / \varepsilon_b) \geq 10^4$ for this example) crosses the abosrption range: $10 < (\varepsilon_I / \varepsilon_b) < 10^4$ the internal force calculated from the GEL2 ST is in good agreement with the external force calculated by external Maxwell ST, (cf. Eq (3b) in this Supplement). The interior of such an extremely abosrbing object is an example of 'GEL system', defined as 'GEL system of $2^{nd}$ kind' in the main paper. So, the question is: what happens in the intermediate abosrption range ($10 < (\varepsilon_I / \varepsilon_b) < 10^4$) between moderate and extreme absorption?

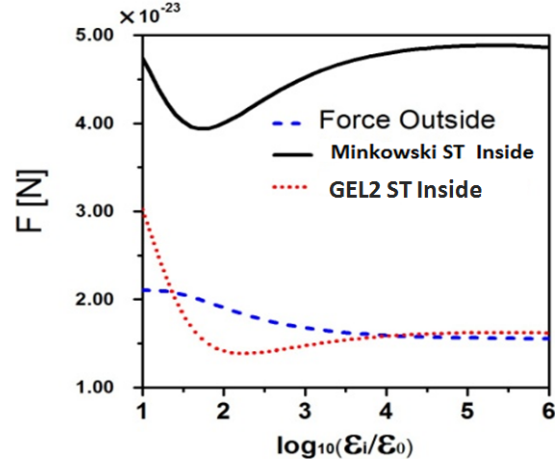

Figure 4s: Illustration of optical momentum transfer from a plane polarized incident plane wave at $\lambda = 1064$ nm with electric vector $E_x = e^{i(kz-\omega t)}$ to a highly absorbing magneto-dielectric sphere with radius $a = 1000$ nm, $\varepsilon_s = 7\varepsilon_0 + i\varepsilon_I$, $\mu_s = 3\mu_0$ embedded in air or vacuum. The time-averaged force outside $\boldsymbol{F}_{out}$ (in Newtons) is calculated by the Minkowski (i.e. Maxwell's) ST, Eq. (3b), on the surface $r=a^+$ using exterior fields. The force $F_{in}$ based on the 'GEL2 ST' (cf. Eq (23) in the main paper), obtained on the surface $r = 0.999999a = a^{--}$ (i.e. extremely close to the object boundary) with interior fields, applies only for extreme losses, where it matches with the force correctly given by Maxwell ST with exterior fields. The Minkowski ST with interior fields (black full line) does not match with the external Maxwell ST. The difference is contained in $\boldsymbol{F}_b$, (cf. ref. [2] and also Eqs (11a) and (13) in supplement S1(c)].

As shown in Fig.4s, in the range $10 < (\varepsilon_I / \varepsilon_b) < 10^4$, in order to determine the final form of the internal Generalized EL ST that supports Eq (9b) of the main paper, the appropriate $\varepsilon_{Eff}$ [cf. Eq (8b) in the main paper] cannot be attributed to certain constitutive parameter values (also cf. Eqs (22a)- (22c) of the main paper). In fact, there is no known ST which yields the total internal force within this absorption range. When $10 < (\varepsilon_I / \varepsilon_b) < 10^4$, only the internal kinetic force on a scatterer (embedded in air) can be calculated by employing Chu ST inside the object. The static or surface force part,

$$\boldsymbol{f}_{Static}^{Chu} = \boldsymbol{f}_{Surface} = \rho_e \boldsymbol{E}_{average} + \rho_m \boldsymbol{H}_{average} = \{\epsilon_o (\boldsymbol{E}_{out} - \boldsymbol{E}_{in}) \cdot \hat{\boldsymbol{n}}\}\left(\frac{\boldsymbol{E}_{out} + \boldsymbol{E}_{in}}{2}\right) + \{\mu_0 (\boldsymbol{H}_{out} - \boldsymbol{H}_{in}) \cdot \hat{\boldsymbol{n}}\}\left(\frac{\boldsymbol{H}_{out} + \boldsymbol{H}_{in}}{2}\right),$$

which arises on the boundary of the scatterer (i.e. at $r = a$) is a hidden quantity. We have discusssed this fact in detail in supplement S1 (a), [see also Eq (2) in this Supplement]. The only way that we have found to calculate the total volume force is to add that hidden quantity of the Chu force manually with the aformentioned internal kinetic force part. But then the question arises: why beyond that specific range $10 < (\varepsilon_I / \varepsilon_b) < 10^4$ the total internal force can be calculated from the GEL2 ST, (cf. Eq (23) in the main paper), without any manual addition of a hidden quantity?

The reason can be understood from the volume force density equation, (Eq (24) in the main paper). For an extremely absorbing dielectric: the GEL2 force equation can be written as:

$\langle f_{\text{GEL2}} \rangle = \frac{1}{2} \text{Re} \left[ (P_{\text{Eff}} \cdot \nabla) E_{\text{in}}^* - i\omega (P_{\text{Eff}} \times B_{\text{in}}^*) \right] = \frac{1}{2} \text{Re} \left[ (i\varepsilon_I E_{\text{in}} \cdot \nabla) E_{\text{in}}^* - i\omega (i\varepsilon_I E_{\text{in}} \times B_{\text{in}}^*) \right]$. So, the value of $P_{\text{Eff}} = (\varepsilon_S - \varepsilon_{\text{Eff}}) E_{\text{in}}$ fixes at $i\varepsilon_I E_{\text{in}}$ beyond the interval $10 < (\varepsilon_I / \varepsilon_b) < 10^4$ of absorption (a plausible and simple explanation is that: as $\text{Re}[\varepsilon_s] \ll \varepsilon_I$, $\text{Re}[\varepsilon_s]$ does not contribute in the $P_{\text{Eff}}$ specifically in the extreme loss range). As a result, the value of $\varepsilon_{\text{Eff}}$ can be recognized as $\text{Re}[\varepsilon_s]$, (cf. Eq (22b) in the main paper) in the cases of extreme absorption when one determines the final form of the GEL ST, (i.e. the GEL2 ST).

However within the absorption range: $10 < (\varepsilon_I / \varepsilon_b) < 10^4$, the surface force $f_{\text{Surface}}$ acts on the boundary of the scatterer due to the significant influence of both $E_{\text{out}}$ and $E_{\text{in}}$ on the surface bound charges: $\rho_e E_{\text{average}}$ or $-(\nabla \cdot P) E_{\text{average}}$ [25]. Notice that $E_{\text{average}}$ contains both the $E_{\text{out}}$ and $E_{\text{in}}$. In consequence, the static force $(P_{\text{Eff}} \cdot \nabla) E_{\text{in}}$, based on the dipole model [26] rather than on the surface bound charge model [25, 26], in the GEL force equation is no longer valid. Such domination of the surface force may also arise in some other unusual situations such as: (i) due to induced whispering gallery modes on the surface of scatterers [27], (ii) due to internal inhomogenity of the scatterers [25], (iii) due to the multiple scattering effects [28]. In such situations the term $(P \cdot \nabla) E_{\text{in}}$ (or $(P_{\text{Eff}} \cdot \nabla) E_{\text{in}}$) in the EL (or GEL) equations, and hence the internal ST of Einstein-Laub (or GEL) does not remain valid. More elaborately, we cannot determine any fixed value for $\varepsilon_{\text{Eff}}$ from the relation: $P_{\text{Eff}} = (\varepsilon_S - \varepsilon_{\text{Eff}}) E_{\text{in}}$ in such cases. Hence, it becomes impossible to determine the final form of the GEL ST, (which is a function of $\varepsilon_{\text{Eff}}$), to determine the total internal force excluding a hidden momentum. In such situations, we do not find other way except employing the hidden force ($f_{\text{Surface}}$) of Chu to get the total force (internal kinetic force plus surface force) of an object embedded in air.

However, beyond the intermediate absorption range ($10 < (\varepsilon_I / \varepsilon_b) < 10^4$), the dominating surface force, that accounts for the effect of both $E_{\text{out}}$ and $E_{\text{in}}$, almost vanishes, as the total force distributes fully with $E_{\text{in}}$ and $H_{\text{in}}$ due to the kinetic force $\langle f_{\text{GEL}} \rangle = \frac{1}{2} \text{Re} \left[ -i\omega (P_{\text{Eff}} \times B_{\text{in}}^*) + i\omega (M_{\text{Eff}} \times D_{\text{in}}^*) \right]$ caused by the conduction currents inside the scatterer [2]. Hence, only for an extremely absorbing object, (i.e. with $(\varepsilon_I / \varepsilon_b) \geq 10^4$), it is possible to determine the total internal force excluding hidden quantities by employing our GEL2 ST, Eq (23) of the main paper.

Now we return to a different discussion topic: the appropriate photon momentum inside such a highly absorbing scatterer embedded in air or vacuum. Conventionally, the photon interaction with carriers (i.e. electrons) is considered associated to the Minkowski photon momentum [6]. But Fig.4s shows that when high losses (i.e. approximately with $(\varepsilon_I / \varepsilon_b) > 10$) occur in a magneto-dielectric object rather than in a dielectric [2,6], the force $F_c$ [2] obtained from the internal Minkowski ST (i.e. with interior wavefields) largely departs from the total force yielded by the exterior Maxwell (or Minkowski) ST (cf. Fig. 4s), since then the bound force $F_b$ [2] is not small [see also Eqs.(11a) and (13) in supplement S1(c)]. Although the interaction of photons and carriers (i.e. electrons) conveys a Minkowski momentum density, the host with any permeability value other than one, significantly contributes to the total force, even when the abosrption of the object is very high such as shown in Fig.4s. On the other hand, neither the EL nor Chu's ST (both supporting the Abrham momentum density) yield the total internal force of such a highly absorbing object (i.e. approximately with $(\varepsilon_I / \varepsilon_b) > 10$) embedded in air. In this connectiion notice that in the Einstein-Balazs thought experiment [15], the box is considered reflectionless or transparent [15,22]; according to our analysis a scatterer even if it produces loss (and a moderate reflection) can be safely modelled as a transparent Einstein-Balazs box up to moderate absorption (i.e. with $(\varepsilon_I / \varepsilon_b) \leq 10$ for several examples in our work) supporting the internal photon momentum as $p_{\text{Abr}}(\text{in})$.

One of the main conclusions in this work is: the *internal non-mechanical momentum density*, *internal ST* and *force* are indeed functions of our defined *effective polarization* and *magnetization* of the object, which widely vary for different systems. However, based on the appropriate effective polarization and magnetization, it is possible to determine the correct internal force on a generic scatterer with the GEL formulations inside a generic object for most of the situations excluding hidden momenta. Still the importance of the Chu type force density equation in [2, 7, 25] cannot be ignored because there may arise some unusual situations [25-28] in which the total volume force cannot be calculated solely from the interior fields excluding hidden quantities. Hence, once again we stress that: 'the dilemma should not be: which force law along with its stress tensor is correct and which one is incorrect?. Rather it should be: *in a specific measurement, which force law along with its ST should be effective and which should not?*'.

## References and Notes

[1] B. A. Kemp, "Resolution of the Abraham-Minkowski debate: Implications for the electromagnetic wave theory of light in matter," J. Appl. Phys. **109**, 111101 (2011).

[2] B. A. Kemp, T.M. Grzegorczyk, and J. A. Kong, "Optical Momentum Transfer to Absorbing Mie Particles," Phys. Rev.Lett. **97**, 133902 (2006).

[3] Y. He, J.-Q. Shen, and S. He, "Consistent formalism for the momentum of electromagnetic waves in lossless dispersive metamaterials and the conservation of momentum," Prog. In Electromag. Research **116**, 81-106 (2011).

[4] R. Loudon, L. Allen and D. F. Nelson, "Propagation of electromagnetic energy and momentum through an absorbing dielectric," Phys. Rev. E **55**, 1071–1085 (1997).

[5] S. M. Barnett and R. Loudon, "On the electromagnetic force on a dielectric medium," J. Phys. B: At. Mol. Opt. Phys. **39**, S671 (2006).

[6] S. M. Barnett and R. Loudon, "The enigma of optical momentum in a medium," Phil. Trans. R. Soc. A **368**, 927 (2010).

[7] B. A. Kemp, T. M. Grzegorczyk and J. A. Kong, "Lorentz Force on Dielectric and Magnetic Particles," Journal of Electromag. Wave Appl. **20**, 827 (2006).

[8] J. D. Jackson, "Classical Electrodynamics," 3rd ed., J. Wiley, New York, 1999. Ch. 6.

[9] D. J. Griffiths, V. Hnizdo, "Mansuripur's paradox," Am. J. of Phys. **81**, 570 (2013).

[10] E. A. Hinds and Stephen M. Barnett, "Momentum exchange between light and a single atom: Abraham or Minkowski?" Phys. Rev. Lett. **102**, 050403 (2009).

[11] Cheng-Wei Qiu, Weiqiang Ding, M.R.C. Mahdy, Dongliang Gao, Tianhang Zhang, Fook Chiong Cheong, Aristide Dogariu, Zheng Wang and Chwee Teck Lim, "Photon momentum transfer in

[12] B.A. Kemp, and T.M. Grzegorczyk, "The observable pressure of light in dielectric fluids", Opt. Lett. **36**, 493 (2011).

[13] M. Nieto-Vesperinas, J.J. Sáenz, R. Gómez-Medina, and L. Chantada, "Optical forces on small magnetodielectric particles," Opt. Express **18**, 11428 (2010).

[14] V. Kajorndejnuku, W. Ding, S. Sukhov, C.-W. Qiu and A.Dogariu, "Linear momentum increase and negative optical forces at dielectric interface," Nat. Photonics **7**, 787 (2013).

[15] P. W. Milonni and R. W. Boyd, "Momentum of Light in a Dielectric Medium," Adv. in Optics and Photonics **2**, 519 (2010).

[16] Note that in Eq (15) and Eq (20), $\boldsymbol{p}_{\text{Photon}}(\text{in}) = m(c/n_s)$ [cf. Eq (14)]. This 'mv' type momentum of photons is considered as the kinetic momentum. The well-known Einstein-Balazs thought experiment [15] leads to $\boldsymbol{p}_{\text{Abr}}(\text{in})$ as the appropriate internal photon momentum, and our analysis support this fact based on the EL ST and force law, [cf. Supplement S0]. However, for an embedded Einstein-Balazs box, although the internal photon momentum is revealed as $\boldsymbol{p}_{\text{Abr}}(\text{in})$, no ST and force law supports it. The possible reason is that some magnitudes of the mechanical momentum in Eq (22) take place inside the box as non-mechanical momenta (as explicitly discussed in supplement S3) and that part can be added with $\boldsymbol{p}_{\text{Abr}}(\text{in})$. Einstein-Balazs box thought experiment always leads to the pure electromagnetic photon momentum inside the box due to the consideration of $\boldsymbol{p}_{\text{Photon}}(\text{in}) = m(c/n_s)$ in Eq (15) and (20). Hence, the thought experiment of Einstein-Balazs' box cannot be a safe procedure to predict the appropriate total non-mechanical momentum of generic objects such as e.g. embedded scatterers.

[17] B. A. Kemp, Comment on "Revisiting the Balazs thought experiment in the presence of loss: electromagnetic-pulse-induced displacement of a positive-index slab having arbitrary complex permittivity and permeability," Appl. Phys. A **110**, 517 (2013).

[18] Maria Dienerowitz, Michael Mazilu, and Kishan Dholakia, "Optical manipulation of nanoparticles: a review," J. Nanophoton. **2**, 021875 (2008)

[19] A simple question that arises is: why two different non-mechanical momentum density formulae [Eqs (27a) and (30b)], along with two different force formulae [Eqs (29) and (30a)], have been shown in Supplement S3 (a) and (b)? The answer is hidden in the single stress tensor formula of Eq (25b). The single ST means that the rate of change of total momentum is constant with time: $\frac{\partial \boldsymbol{p}_{\text{Total}}}{\partial t} = \nabla \cdot \bar{\bar{\text{T}}} = \boldsymbol{f} + \frac{\partial}{\partial t}\boldsymbol{G}$. But at different instants, the force density $\boldsymbol{f}(t)$ and the non-mechanical momentum density $\boldsymbol{G}(t)$ may not be expressed by unique equations. At different instants the mechanical and non-mechanical momenta take on arbitrary expressions, rather than a fixed expression of Eq (22) and (23), respectively. In consequence, a plausible approach to handle such complex situations is to consider the time averaged formulae and to test different situations by equations: (i) $\langle \boldsymbol{F}_{\text{Total}} \rangle(\text{in}) = \oint \langle \bar{\bar{T}}^{\text{in}} \rangle \cdot d\boldsymbol{s} = \int \langle \boldsymbol{f}^{\text{in}} \rangle dv.$ and (ii) $\langle \boldsymbol{F}_{\text{Total}} \rangle(\text{out}) \approx \langle \boldsymbol{F}_{\text{Total}} \rangle(\text{in})$. According to these two tests, Eq (30a) is the appropriate force law, which can even handle the internal force in 'kinetic systems', too. As a result, the more appropriate non-mechanical momentum density should be Eq (30b) that leads to the time-varying version of force Eq (30a) and also does not require any hidden momentum.

[20] K.T.Mcdonald, "Biot-Savart versus Einstein-Laub Force Law," Joseph Henry Laboratories, Princeton University, Princeton, NJ 08544, (2013). http://www.physics.princeton.edu/~mcdonald/examples/laub.pdf.

[21] We can consider the case of an embedded object (*generalized Einstein Laub* system of first kind) discussed in Supplement S2. Only for this case the parameters $\varepsilon_{Eff}$ and $\mu_{Eff}$ can be written with exact values: $\varepsilon_b$ and $\mu_b$ respectively. Thus Eq (27b) can be expressed as:

$$G_{GEL}^{Transition} = G_{REL}^{Transition} = \varepsilon_b \mu_b (E_{in} \times H_{in}) = [G_{Material} + G_{Electromag}] = [\varepsilon_b \mu_b - 1/c^2](E_{in} \times H_{in}) + (E_{in} \times H_{in})/c^2 \quad \text{(N21a)}.$$

Now, Eq (30b) can also be written as:

$$G_{GEL1} = [G_{Material} + G_{Electromag}] = [(\varepsilon_b \mu_b - 1/c^2)(E_{in} \times H_{in}) - (P_{Eff} \times M_{Eff})] + (E_{in} \times H_{in})/c^2 \quad \text{(N21b)}.$$

GEL1 means the first kind Generalized Einstein-Laub formulation, applicable only for the '*generalized Einstein Laub* system of first kind'. According to Eqs (N21a) and (N21b), $p_{Abr}(in)$ can be considered as the pure electromagnetic part of total non-mechanical photon momentum, which maintains a fixed expression inside an embedded object. In contrast, the 'Material Induced Momentum' term, $p_{Material}(in)$, connected with $G_{Material}$ in Eqs. (N21a) and (N21b), may achieve different expressions (i.e. magnitudes) at different instants [19]. The two thought experiments of Case-1 and Case-2 in Supplement S2 lead only to that fixed expression of the pure electromagnetic part $p_{Abr}(in)$ rather than to the total non-mechanical momentum [19]. Note that for a 'kinetic system' $p_{Material}(in)$, defined in Eqs (27c) is zero. This is the reason why in the well-known Einstein-Balazs thought experiment [15] the non-mechanical photon momentum reveals as the one of Abraham, $p_{Abr}(in)$.

[22] S.M. Barnett, "Resolution of the Abraham-Minkowski Dilemma," Phys. Rev. Lett. **104**, 070401 (2010).
[23] B. A. Kemp, T. M. Grzegorczyk, and J. A. Kong, "Ab initio study of the radiation pressure on dielectric and magnetic media," Opt. Express **13**, 9280 (2005).
[24] Winston Frias and Andrei I. Smolyakov, "Electromagnetic forces and internal stresses in dielectric media," Phys. Rev. E **85**, 046606 (2012).
[25] H, Chen, B, Zhang, Y, Luo, B, A. Kemp, J, Zhang, L, Ran, and B. I. Wu, "Lorentz force and radiation pressure on a spherical cloak," Phys. Rev. A **80,** 011808 ( 2009).
[26] M. Mansuripur and A. R. Zakharian, "Maxwell's macroscopic equations, the energy-momentum postulates, and the Lorentz law of force," Phys. Rev. E **79**, 026608 (2009).
[27] F. J. V.Valero and M. Nieto-Vesperinas , "Optical forces on cylinders near subwavelength slits: effects of extraordinary transmission and excitation of Mie resonances," Opt.Express, 20, 13368 (2012).
[28] T. M. Grzegorczyk, B. A. Kemp, and J. A. Kong, "Stable Optical Trapping Based on Optical Binding Forces," Phys. Rev. Lett. **96**, 113903 (2006).